\documentclass[10pt]{article}
\usepackage{graphicx}
\usepackage{amsmath}
\usepackage{amssymb}
\usepackage{caption2}
\begin{document}
\bibliographystyle{prsty}
\begin{center}
{\large {\bf \sc{  Reanalysis of  the $X(4140)$  as   axialvector tetraquark state with  QCD sum rules }}} \\[2mm]
Zhi-Gang  Wang \footnote{E-mail: zgwang@aliyun.com.  }   \\
 Department of Physics, North China Electric Power University, Baoding 071003, P. R. China
\end{center}

\begin{abstract}
In this article, we take the $X(4140)$ as the diquark-antidiquark type   $cs\bar{c}\bar{s}$ tetraquark state with $J^{PC}=1^{++}$,  and  study the mass and pole residue with the QCD sum rules in details by constructing two types interpolating currents. The numerical results $M_{X_{L,+}}=3.95\pm0.09\,\rm{GeV}$ and $M_{X_{H,+}}=5.00\pm0.10\,\rm{GeV}$
disfavor assigning the $X(4140)$ to be the $J^{PC}=1^{++}$ diquark-antidiquark type $cs\bar{c}\bar{s}$  tetraquark state. Moreover, we obtain the masses of the   $J^{PC}=1^{+-}$ diquark-antidiquark type $cs\bar{c}\bar{s}$ tetraquark states as a byproduct. The present predictions can be confronted to the experimental data in the future.
\end{abstract}



\section{Introduction}

In 2009,   the CDF collaboration observed the $X(4140)$ for the first time  in the $J/\psi\phi$ invariant mass distribution  in the exclusive $B^+ \to J/\psi\,\phi K^+$ decays in $p\bar{p}$ collisions at $\sqrt{s} = 1.96\,\rm{ TeV}$ with a statistical
significance  more than $3.8 \sigma$  \cite{CDF0903}.
 In 2011, the CDF collaboration confirmed the
$X(4140)$ in the $B^\pm\rightarrow J/\psi\,\phi K^\pm$ decays  with
 a statistical significance more  than $5\sigma$, and  observed an evidence for the  $X(4274)$ in the $J/\psi\phi$ invariant mass distribution  with a statistical significance about   $3.1\sigma$
\cite{CDF1101}.
In 2013, the CMS collaboration   confirmed the $X(4140)$  in the $B^\pm \to J/\psi \phi K^\pm$ decays in $pp$ collisions at $\sqrt{s} = 7\,\rm{ TeV}$ collected with the CMS detector at the LHC with a statistical significance more than $5 \sigma$
 \cite{CMS1309},  the D0 collaboration confirmed  the  $X(4140)$  in the  $B^+ \to J/\psi \phi K^+$ decays with a statistical significance of $3.1\sigma$
based on the data sample corresponds to an integrated luminosity of $10.4\, \rm{fb}^{-1}$  of   $p\bar{p}$ collisions at $\sqrt{s} = 1.96\,\rm{ TeV}$
 \cite{D0-1309}. There have been several possible assignments for the $X(4140)$ since its first  observation  by the CDF collaboration \cite{CDF0903}, such as a molecular
state \cite{X4140-Molecule}, a tetraquark state \cite{X4140-tetraquark-Stancu,X4140-tetraquark-Drenska,X4140-tetraquark,X4140-Wang1502,X4140-tetraquark-Lebed}, a hybrid state \cite{X4140-hybrid, X4140-hybrid-Wang0903} or a rescattering effect \cite{X4140-rescat}.

Recently, the LHCb collaboration performed the first full amplitude analysis of the decays $B^+\to J/\psi \phi K^+$ with $J/\psi\to\mu^+\mu^-$, $\phi\to K^+K^-$   with a data sample corresponds to an integrated luminosity of $3\rm{fb}^{-1}$ of $pp$ collision data collected at $\sqrt{s}=7$ and $8$ TeV with the LHCb detector, and observed  that  the data cannot be described by a model that contains only excited kaon states decaying into $\phi K^+$ \cite{LHCb-4500-1606.07895,LHCb-4500-1606.07898}.
The LHCb collaboration confirmed the two old particles $X(4140)$ and $X(4274)$ in the $J/\psi \phi$ invariant mass distributions  with statistical significances $8.4\sigma$ and $6.0\sigma$, respectively,  and determined the    spin-parity-change-conjugation  to be $J^{PC} =1^{++}$ with statistical significances $5.7\sigma$ and $5.8\sigma$, respectively \cite{LHCb-4500-1606.07895,LHCb-4500-1606.07898}. Moreover, LHCb collaboration observed  the two new particles $X(4500)$ and $X(4700)$ in the $J/\psi \phi$ invariant mass distributions with statistical significances $6.1\sigma$ and $5.6\sigma$, respectively,  and determined the    spin-parity-change-conjugation to be $J^{PC} =0^{++}$ with statistical significances $4.0\sigma$ and $4.5\sigma$, respectively \cite{LHCb-4500-1606.07895,LHCb-4500-1606.07898}. The measured Breit-Wigner masses and widths are
\begin{flalign}
 & X(4140) : M = 4146.5 \pm 4.5 ^{+4.6}_{-2.8} \mbox{ MeV}\, , \, \Gamma = 83 \pm 21 ^{+21}_{-14} \mbox{ MeV} \, , \nonumber\\
 & X(4274) : M = 4273.3 \pm 8.3 ^{+17.2}_{-3.6} \mbox{ MeV}\, , \, \Gamma = 56 \pm 11 ^{+8}_{-11} \mbox{ MeV} \, ,\nonumber \\
 & X(4500) : M = 4506 \pm 11 ^{+12}_{-15} \mbox{ MeV} \, ,\, \Gamma = 92 \pm 21 ^{+21}_{-20} \mbox{ MeV} \, , \nonumber\\
 & X(4700) : M = 4704 \pm 10 ^{+14}_{-24} \mbox{ MeV} \, ,\, \Gamma = 120 \pm 31 ^{+42}_{-33} \mbox{ MeV} \, .
\end{flalign}

The LHCb collaboration determined the quantum numbers of the $X(4140)$ to be $J^{PC}=1^{++}$, which rules  out the $0^{++}$ or $2^{++}$ $D_s^{*+}D_s^{*-}$ molecule assignment.  In the constituent diquark model, the masses of the ground state $cs\bar{c}\bar{s}$ tetraquark states with  $J^{PC}=0^{-+}$, $1^{-+}$ are about $4.3\,\rm{GeV}$ \cite{X4140-tetraquark-Drenska}, while the  masses of the ground state  diquark-antidiquark type $cs\bar{c}\bar{s}$ tetraquark states with $J^{PC}=0^{++}$ and $2^{++}$ from the QCD sum rules are about $3.98\pm0.08\,\rm{GeV}$ and $4.13\pm0.08\,\rm{GeV}$, respectively \cite{X4140-Wang1502}. In Ref.\cite{X4140-tetraquark-Lebed},  Lebed and Polosa  propose that the  $X(3915)$ is the ground state scalar-diquark-scalar-antidiquark type
scalar $cs\bar{c} \bar{s}$ tetraquark state  according to  lacking of the observed
decays to the final states $D\bar D$ and $D^*\bar{D}^*$, and attribute  the only known decay to the final state $J/\psi \omega$ to the $\omega-\phi$ mixing effect. In Ref.\cite{Wang-4500}, we tentatively assign the  $X(3915)$ and $X(4500)$ to be the ground state and the first radial excited state of the axialvector-diquark-axialvector-antidiquark  type scalar $cs\bar{c}\bar{s}$ tetraquark states, respectively,  and study their  masses and pole residues in details with the QCD sum rules, and obtain the values,
 \begin{eqnarray}
M_{X(3915)}&=&3.92^{+0.19}_{-0.18}\,\rm{GeV} \, ,  \,\,\, {\rm Experimental\,\, value} \,\,\,\,3918.4\pm 1.9\,\rm{ MeV} \, \cite{PDG}\,   , \nonumber\\
M_{X(4500)}&=&4.50^{+0.08}_{-0.09}\,\rm{GeV} \, ,  \,\,\, {\rm Experimental\,\, value} \,\,\,\,4506 \pm 11 ^{+12}_{-15} \, \rm {MeV}\, \cite{LHCb-4500-1606.07895,LHCb-4500-1606.07898}\,   ,
\end{eqnarray}
which are consistent with the experimental data. The inclusion of the  first radial excited state beyond the ground state in the QCD sum rules leads to smaller ground state mass \cite{X4140-Wang1502}, which happens to lie in the same energy region of the $X(3915)$. If the masses of the  ground state diquark-antidiquark  type $0^{++}$ and $2^{++}$ $cs\bar{c}\bar{s}$ tetraquark states are about $3.9\,\rm{GeV}$ and $4.1\,\rm{GeV}$, respectively, we would expect that the ground state diquark-antidiquark  type $1^{++}$ $cs\bar{c}\bar{s}$ tetraquark state has the mass about $3.9-4.1\,\rm{GeV}$.

In Ref.\cite{X4140-tetraquark-Stancu}, F. Stancu calculates  the mass spectrum of the $c\bar{c}s\bar{s}$ tetraquark states   within a simple quark model with chromomagnetic interaction and effective quark masses extracted from meson and baryon spectra, and obtain the two lowest masses $4195\,\rm{MeV}$ and $4356\,\rm{MeV}$ of  the tetraquark states  with $J^{PC}=1^{++}$. The value $4195\,\rm{MeV}$ is consistent with the experimental data $4146.5 \pm 4.5 ^{+4.6}_{-2.8} \mbox{ MeV}$. In the simple chromomagnetic interaction model, there are no correlated quarks or diquarks \cite{X4140-tetraquark-Stancu}.

The scattering amplitude for one-gluon exchange is proportional to
\begin{eqnarray}
T^a_{ki}
T^a_{lj}&=&-\frac{1}{3}(\delta_{jk}\delta_{il}-\delta_{ik}\delta_{jl})
 +\frac{1}{6}(\delta_{jk}\delta_{il}+\delta_{ik}\delta_{jl})\, ,
\end{eqnarray}
where the $T^a$ is the generator of the $SU_c(3)$ gauge group, and the $i,j$
and $k,l$ are the color indexes of the two quarks in the incoming
and outgoing channels respectively.   The negative sign
in front of the antisymmetric  antitriplet indicates the interaction
is attractive, which favors the formation of diquark states in the color
 antitriplet \cite{One-gluon}, so we usually take the diquarks in  color
 antitriplet as the basic constituents in studying the baryon states, tetraquark states and pentaquark states.
The diquarks $\varepsilon^{ijk} q^{T}_j C\Gamma q^{\prime}_k$ in color
 antitriplet have  five  structures  in Dirac spinor space, where the $i$, $j$ and $k$ are color indexes, $C\Gamma=C\gamma_5$, $C$, $C\gamma_\mu \gamma_5$,  $C\gamma_\mu $ and $C\sigma_{\mu\nu}$ for the scalar, pseudoscalar, vector, axialvector  and  tensor diquarks, respectively.
The  stable diquark configurations are the scalar ($C\gamma_5$) and axialvector ($C\gamma_\mu$) diquark states from the QCD sum rules \cite{WangDiquark,WangLDiquark}, we can construct the tetraquark states using the scalar  or axialvector  diquarks rather than the uncorrelated quarks to obtain the lowest masses.

In Ref.\cite{WangHuangTao-3900},  we  study the masses and pole residues of the axialvector hidden-charm tetraquark states in details with the QCD sum rules, and observe
that the predictions $M_{X(3872)}=3.87^{+0.09}_{-0.09}\,\rm{GeV}$ and $M_{Z_c(3900)}=3.91^{+0.11}_{-0.09}\,\rm{GeV}$ support assigning    the $X(3872)$ and $Z_c(3900)$  to be the $1^{++}$
and $1^{+-}$ diquark-antidiquark type tetraquark states, respectively. If we take the $X(4140)$ as the hidden-strange cousin of the $X(3872)$, then the mass difference  $M_{X(4140)}-M_{X(3872)}=275\,\rm{MeV}$, the $SU(3)$ breaking effect is about $m_s-m_q=135\,\rm{MeV}$, which is consistent with our naive expectation.
In Ref.\cite{Chen-Zhu-2011}, Chen and Zhu obtain the value $4.07\pm 0.10\,\rm{GeV}$ for the mass of the $C\gamma_5 \otimes \gamma_\mu C+C\gamma_\mu \otimes \gamma_5 C$ type $cs\bar{c}\bar{s}$ tetraquark states based on the QCD sum rules, the theoretical value $4.07\pm 0.10\,\rm{GeV}$ overlaps with the experimental value $4146.5 \pm 4.5 ^{+4.6}_{-2.8} \mbox{ MeV}$, which supports assigning the $X(4140)$ to be the axialvector tetraquark state \cite{Chen1606}. Although the masses of the axialvector tetraquark states are calculated with the QCD sum rules, the routines  are  different \cite{WangHuangTao-3900,Chen-Zhu-2011}. In Ref.\cite{WangHuangTao-3900}, we study the energy scale dependence of the QCD spectral densities for the first time, and in subsequent works  \cite{WangTetraquarkCTP,Wang-4660-2014,Wang-Huang-NPA-2014},  we suggest an empirical energy scale formula,
\begin{eqnarray}
\mu&=&\sqrt{M^2_{X/Y/Z}-(2{\mathbb{M}}_Q)^2} \, ,
 \end{eqnarray}
 with the effective heavy quark masses ${\mathbb{M}}_Q$ to determine the ideal energy scales of the QCD spectral densities of the hidden-charm and the hidden-bottom tetraquark states  in the QCD sum rules.

Before the work \cite{Chen-Zhu-2011},  we performed a systematic study of the mass spectrum of the axialvector hidden-charm and hidden-bottom tetraquark states using the QCD sum rules, and obtained the ground state masses $M_{cq\bar{c}\bar{q}}=4.32\pm0.18\,\rm{GeV}$ and  $M_{cs\bar{c}\bar{s}}=4.40\pm0.16\,\rm{GeV}$ \cite{Wang-Axial-V-tetraquark}, the mass breaking effect $M_{cs\bar{c}\bar{s}}-M_{cq\bar{c}\bar{q}}=80\,\rm{MeV}$, which is much smaller than the experimental value $M_{X(4140)}-M_{X(3872)}=275\,\rm{MeV}$. In Ref.\cite{Wang-Axial-V-tetraquark}, we extract the masses from the QCD spectral densities at the energy scale $\mu=1\,\rm{GeV}$, which is much smaller than the optimal energy scales determined by the empirical energy scale formula, and results in much larger mass  $M_{cq\bar{c}\bar{q}}=4.32\pm0.18\,\rm{GeV}$ compared to the mass  $M_{X(3872)/Z_c(3900)}\approx 3.9\,\rm{GeV}$ extracted at the optimal energy scales \cite{WangHuangTao-3900}.

In Ref.\cite{Wang-4660-2014}, we  study the masses and pole residues of the $J^{PC}=1^{-\pm}$ hidden-charm tetraquark states at the optimal energy scales  with the QCD sum rules.   The predicted masses of the  tetraquark states with symbolic quark structures $c\bar{c}s\bar{s}$ and $c\bar{c}(u\bar{u}+d\bar{d})/\sqrt{2}$ support  assigning the $Y(4660)$ to  be  the $1^{--}$ diquark-antidiquark type tetraquark state, the mass difference
 $M_{c\bar{c}s\bar{s}}-M_{c\bar{c}(u\bar{u}+d\bar{d})/\sqrt{2}}=40\,\rm{MeV}$ is even smaller compared to the value $80\,\rm{MeV}$ obtained in Ref.\cite{Wang-Axial-V-tetraquark}.

 Now we can draw the conclusion tentatively that the QCD sum rules support  smaller $SU(3)$ breaking effect than our  naive expectation. It is interesting to perform detailed studies of the $X(4140)$ as the axialvector $cs\bar{c}\bar{s}$ tetraquark state based on the QCD  sum rules.

In this article, we take the $X(4140)$ as the axialvector $cs\bar{c}\bar{s}$ tetraquark state, construct the diquark-antidiquark type axialvector currents, calculate the contributions of the vacuum condensates up to dimension 10 in the operator product expansion in a  consistent way, use the empirical energy scale formula to determine the ideal energy scales of the QCD spectral densities, and  study the ground state masses and pole residues in details with the QCD sum rules. We want to obtain additional support in assigning the $X(4140)$ to be the $1^{++}$ $cs\bar{c}\bar{s}$ tetraquark state   from  the QCD sum rules.

The article is arranged as follows:  we derive the QCD sum rules for the masses and pole residues of  the axialvector $cs\bar{c}\bar{s}$ tetraquark states in section 2; in section 3, we present the numerical results and discussions; section 4 is reserved for our conclusion.

\section{QCD sum rules for  the  axialvector $cs\bar{c}\bar{s}$ tetraquark states  }
In the following, we write down  the two-point correlation functions $\Pi^{\pm}_{\mu\nu}(p)$  in the QCD sum rules,
\begin{eqnarray}
\Pi_{\mu\nu}^{\pm}(p)&=&i\int d^4x e^{ip \cdot x} \langle0|T\left\{J^{\pm}_\mu(x){J_\nu^{\pm}}^{\dagger}(0)\right\}|0\rangle \, ,
\end{eqnarray}
where $J_\mu^{\pm}(x)=J^{L,\pm}_\mu(x)$, $J^{H,\pm}_\mu(x)$,
\begin{eqnarray}
J^{L,\pm}_\mu(x)&=&\frac{\epsilon^{ijk}\epsilon^{imn}}{\sqrt{2}}\left\{s^j(x)C\gamma_5c^k(x) \bar{s}^m(x)\gamma_\mu C \bar{c}^n(x)\pm s^j(x)C\gamma_\mu c^k(x)\bar{s}^m(x)\gamma_5C \bar{c}^n(x) \right\} \, , \\
J^{H,\mp}_\mu(x)&=&\frac{\epsilon^{ijk}\epsilon^{imn}}{\sqrt{2}}\left\{s^j(x)Cc^k(x) \bar{s}^m(x)\gamma_5\gamma_\mu C \bar{c}^n(x)\pm s^j(x)C\gamma_\mu \gamma_5 c^k(x)\bar{s}^m(x)C \bar{c}^n(x) \right\} \, ,
\end{eqnarray}
 the $i$, $j$, $k$, $m$, $n$ are color indexes, the $C$ is the charge conjunction matrix. We choose  the   currents $J^{L/H,+}_\mu(x)$ to interpolate the
  $J^{PC}=1^{++}$ diquark-antidiquark type hidden-charm tetraquark states.  Under charge conjunction transform $\widehat{C}$, the currents $J^{L/H,\pm}_\mu(x)$ have the properties,
\begin{eqnarray}
\widehat{C}J^{L,\pm}_{\mu}(x)\widehat{C}^{-1}&=& \pm J^{L,\pm}_\mu(x)\, , \nonumber\\
\widehat{C}J^{H,\mp}_{\mu}(x)\widehat{C}^{-1}&=& \mp J^{H,\mp}_\mu(x)\, ,
\end{eqnarray}
which originate from the charge conjunction properties of the scalar, pseudoscalar, axialvector and vector diquark states,
\begin{eqnarray}
\widehat{C}\left[\epsilon^{ijk}q^j C\gamma_5 c^k\right]\widehat{C}^{-1}&=&\epsilon^{ijk}\bar{q}^j \gamma_5 C \bar{c}^k \, , \nonumber\\
\widehat{C}\left[\epsilon^{ijk}q^j C  c^k\right]\widehat{C}^{-1}&=&\epsilon^{ijk}\bar{q}^j   C \bar{c}^k \, , \nonumber\\
\widehat{C}\left[\epsilon^{ijk}q^j C\gamma_\mu c^k\right]\widehat{C}^{-1}&=&\epsilon^{ijk}\bar{q}^j \gamma_\mu C \bar{c}^k \, ,\nonumber\\
\widehat{C}\left[\epsilon^{ijk}q^j C\gamma_\mu \gamma_5 c^k\right]\widehat{C}^{-1}&=&-\epsilon^{ijk}\bar{q}^j \gamma_5\gamma_\mu C \bar{c}^k \, ,
\end{eqnarray}
where $q=u,d,s$. Naively, we expect that the currents $J^{H,\pm}_\mu(x)$ couple  to the hidden-charm tetraquark states with higher masses than that of  the currents $J^{L,\pm}_\mu(x)$, as  the  scalar ($C\gamma_5$) and axialvector ($C\gamma_\mu$) diquark states are much stable compared to  the corresponding  pseudoscalar ($C$) and vector ($C\gamma_\mu\gamma_5$) diquark states \cite{WangDiquark,WangLDiquark}.
In this article, we  study the   $J^{PC}=1^{+-}$ diquark-antidiquark type hidden-charm tetraquark states
as a byproduct.

At the phenomenological side,  we insert  a complete set of intermediate hadronic states with
the same quantum numbers as the current operators $J^{L/H,\pm}_\mu(x)$ into the
correlation functions $\Pi^{\pm}_{\mu\nu}(p)$  to obtain the hadronic representation
\cite{SVZ79,Reinders85}. After isolating the ground state hidden-charm  tetraquark states $X_{L/H,\pm}$ and $X^{\prime}_{L/H,\pm}$
contributions  from the pole terms, we get the following result,
\begin{eqnarray}
\Pi^{\pm}_{\mu\nu}(p)&=&\frac{\lambda_{X_{L/H,\pm}}^2}{M_{X_{L/H,\pm}}^2-p^2}\left(-g_{\mu\nu} +\frac{p_\mu p_\nu}{p^2}\right) +\frac{\widetilde{\lambda}_{X_{L/H,\pm}}^2}{\widetilde{M}_{X_{L/H,\pm}}^2-p^2}\,p_\mu p_\nu+\cdots \, \, , \nonumber\\
&=&\Pi_{L/H,\pm}(p)\left(-g_{\mu\nu} +\frac{p_\mu p_\nu}{p^2}\right) +\widetilde{\Pi}_{L/H,\pm}(p)\,p_\mu p_\nu \, \, ,
\end{eqnarray}
where the pole residues (or coupling constants) $\lambda_{X_{L/H,\pm}}$ and $\widetilde{\lambda}_{X_{L/H,\pm}}$ are defined by
\begin{eqnarray}
 \langle 0|J^{L/H,\pm}_\mu(0)|X_{L/H,\pm}(p)\rangle=\lambda_{X_{L/H,\pm}} \, \varepsilon_\mu \, , \nonumber\\
 \langle 0|J^{L/H,\pm}_\mu(0)|X^{\prime}_{L/H,\pm}(p)\rangle=\widetilde{\lambda}_{X_{L/H,\pm}} \, p_\mu \, ,
\end{eqnarray}
the $\varepsilon_\mu$ are the polarization vectors of the axialvector tetraquark states $X_{L/H,\pm}$. In this article, we choose the tensor structure $-g_{\mu\nu} +\frac{p_\mu p_\nu}{p^2}$ for analysis, the pseudoscalar tetraquark states $X^{\prime}_{L/H,\pm}$ have no contaminations.

 In the following,  we briefly outline  the operator product expansion for the correlation functions $\Pi^{\pm}_{\mu\nu}(p)$  in perturbative
QCD.  We contract the quark fields in the correlation functions
$\Pi^{\pm}_{\mu\nu}(p)$ with Wick theorem firstly, and obtain the results:
\begin{eqnarray}
\Pi^{L,\pm}_{\mu\nu}(p)&=&-\frac{i\epsilon^{ijk}\epsilon^{imn}\epsilon^{i^{\prime}j^{\prime}k^{\prime}}\epsilon^{i^{\prime}m^{\prime}n^{\prime}}}{2}\int d^4x e^{ip \cdot x}   \nonumber\\
&&\left\{{\rm Tr}\left[ \gamma_5C^{kk^{\prime}}(x)\gamma_5 CS^{jj^{\prime}T}(x)C\right] {\rm Tr}\left[ \gamma_\nu C^{n^{\prime}n}(-x)\gamma_\mu C S^{m^{\prime}mT}(-x)C\right] \right. \nonumber\\
&&+{\rm Tr}\left[ \gamma_\mu C^{kk^{\prime}}(x)\gamma_\nu CS^{jj^{\prime}T}(x)C\right] {\rm Tr}\left[ \gamma_5 C^{n^{\prime}n}(-x)\gamma_5 C S^{m^{\prime}mT}(-x)C\right] \nonumber\\
&&-t\,{\rm Tr}\left[ \gamma_\mu C^{kk^{\prime}}(x)\gamma_5 CS^{jj^{\prime}T}(x)C\right] {\rm Tr}\left[ \gamma_\nu C^{n^{\prime}n}(-x)\gamma_5 C S^{m^{\prime}mT}(-x)C\right] \nonumber\\
 &&\left.-t\,{\rm Tr}\left[ \gamma_5 C^{kk^{\prime}}(x)\gamma_\nu CS^{jj^{\prime}T}(x)C\right] {\rm Tr}\left[ \gamma_5 C^{n^{\prime}n}(-x)\gamma_\mu C S^{m^{\prime}mT}(-x)C\right] \right\} \, ,
\end{eqnarray}

\begin{eqnarray}
\Pi^{H,\mp}_{\mu\nu}(p)&=&-\frac{i\epsilon^{ijk}\epsilon^{imn}\epsilon^{i^{\prime}j^{\prime}k^{\prime}}\epsilon^{i^{\prime}m^{\prime}n^{\prime}}}{2}\int d^4x e^{ip \cdot x}   \nonumber\\
&&\left\{{\rm Tr}\left[ \gamma_5\overline{C}^{kk^{\prime}}(x)\gamma_5 CS^{jj^{\prime}T}(x)C\right] {\rm Tr}\left[ \gamma_\nu \overline{C}^{n^{\prime}n}(-x)\gamma_\mu C S^{m^{\prime}mT}(-x)C\right] \right. \nonumber\\
&&+{\rm Tr}\left[ \gamma_\mu \overline{C}^{kk^{\prime}}(x)\gamma_\nu CS^{jj^{\prime}T}(x)C\right] {\rm Tr}\left[ \gamma_5 \overline{C}^{n^{\prime}n}(-x)\gamma_5 C S^{m^{\prime}mT}(-x)C\right] \nonumber\\
&&+t\,{\rm Tr}\left[ \gamma_\mu \overline{C}^{kk^{\prime}}(x)\gamma_5 CS^{jj^{\prime}T}(x)C\right] {\rm Tr}\left[ \gamma_\nu \overline{C}^{n^{\prime}n}(-x)\gamma_5 C S^{m^{\prime}mT}(-x)C\right] \nonumber\\
 &&\left.+t\,{\rm Tr}\left[ \gamma_5 \overline{C}^{kk^{\prime}}(x)\gamma_\nu CS^{jj^{\prime}T}(x)C\right] {\rm Tr}\left[ \gamma_5 \overline{C}^{n^{\prime}n}(-x)\gamma_\mu C S^{m^{\prime}mT}(-x)C\right] \right\} \, ,
\end{eqnarray}
where $t=\pm$, $\overline{C}_{ij}(x)=\gamma_5 C_{ij}(x)\gamma_5$,
 the $S_{ij}(x)$ and $C_{ij}(x)$ are the full  $s$ and $c$ quark propagators, respectively \cite{Reinders85,Pascual-1984},
\begin{eqnarray}
S^{ij}(x)&=& \frac{i\delta_{ij}\!\not\!{x}}{ 2\pi^2x^4}
-\frac{\delta_{ij}m_s}{4\pi^2x^2}-\frac{\delta_{ij}\langle
\bar{s}s\rangle}{12} +\frac{i\delta_{ij}\!\not\!{x}m_s
\langle\bar{s}s\rangle}{48}-\frac{\delta_{ij}x^2\langle \bar{s}g_s\sigma Gs\rangle}{192}+\frac{i\delta_{ij}x^2\!\not\!{x} m_s\langle \bar{s}g_s\sigma
 Gs\rangle }{1152}\nonumber\\
&& -\frac{ig_s G^{a}_{\alpha\beta}t^a_{ij}(\!\not\!{x}
\sigma^{\alpha\beta}+\sigma^{\alpha\beta} \!\not\!{x})}{32\pi^2x^2} -\frac{i\delta_{ij}x^2\!\not\!{x}g_s^2\langle \bar{s} s\rangle^2}{7776} -\frac{\delta_{ij}x^4\langle \bar{s}s \rangle\langle g_s^2 GG\rangle}{27648}-\frac{1}{8}\langle\bar{s}_j\sigma^{\mu\nu}s_i \rangle \sigma_{\mu\nu} \nonumber\\
&&   -\frac{1}{4}\langle\bar{s}_j\gamma^{\mu}s_i\rangle \gamma_{\mu }+\cdots \, ,
\end{eqnarray}
\begin{eqnarray}
C_{ij}(x)&=&\frac{i}{(2\pi)^4}\int d^4k e^{-ik \cdot x} \left\{
\frac{\delta_{ij}}{\!\not\!{k}-m_c}
-\frac{g_sG^n_{\alpha\beta}t^n_{ij}}{4}\frac{\sigma^{\alpha\beta}(\!\not\!{k}+m_c)+(\!\not\!{k}+m_c)
\sigma^{\alpha\beta}}{(k^2-m_c^2)^2}\right.\nonumber\\
&&\left. +\frac{g_s D_\alpha G^n_{\beta\lambda}t^n_{ij}(f^{\lambda\beta\alpha}+f^{\lambda\alpha\beta}) }{3(k^2-m_c^2)^4}
-\frac{g_s^2 (t^at^b)_{ij} G^a_{\alpha\beta}G^b_{\mu\nu}(f^{\alpha\beta\mu\nu}+f^{\alpha\mu\beta\nu}+f^{\alpha\mu\nu\beta}) }{4(k^2-m_c^2)^5}+\cdots\right\} \, , \nonumber \\
\end{eqnarray}
\begin{eqnarray}
f^{\lambda\alpha\beta}&=&(\!\not\!{k}+m_c)\gamma^\lambda(\!\not\!{k}+m_c)\gamma^\alpha(\!\not\!{k}+m_c)\gamma^\beta(\!\not\!{k}+m_c)\, ,\nonumber\\
f^{\alpha\beta\mu\nu}&=&(\!\not\!{k}+m_c)\gamma^\alpha(\!\not\!{k}+m_c)\gamma^\beta(\!\not\!{k}+m_c)\gamma^\mu(\!\not\!{k}+m_c)\gamma^\nu(\!\not\!{k}+m_c)\, ,
\end{eqnarray}
and  $t^n=\frac{\lambda^n}{2}$, the $\lambda^n$ is the Gell-Mann matrix,  $D_\alpha=\partial_\alpha-ig_sG^n_\alpha t^n$ \cite{Reinders85}, we add the superscripts $L$ and $H$ to denote which interpolating current is used.
Then we compute  the integrals both in the coordinate space and in the momentum space,  and obtain the correlation functions $\Pi^{L/H,\pm}_{\mu\nu}(p)$
at the quark level. The calculations are   straightforward  but  tedious.
Once the analytical expressions of the correlation functions $\Pi_{L/H,\pm}(p)$ are gotten, we can obtain the QCD spectral densities $\rho_{L/H,\pm}(s)$ through   dispersion  relation.
In Eq.(14), we retain the terms $\langle\bar{s}_j\sigma_{\mu\nu}s_i \rangle$ and $\langle\bar{s}_j\gamma_{\mu}s_i\rangle$ originate from the Fierz re-ordering of the $\langle s_i \bar{s}_j\rangle$ to  absorb the gluons  emitted from the heavy quark lines to form $\langle\bar{s}_j g_s G^a_{\alpha\beta} t^a_{mn}\sigma_{\mu\nu} s_i \rangle$ and $\langle\bar{s}_j\gamma_{\mu}s_ig_s D_\nu G^a_{\alpha\beta}t^a_{mn}\rangle$   to extract the mixed condensate and four-quark condensates $\langle\bar{s}g_s\sigma G s\rangle$ and $g_s^2\langle\bar{s}s\rangle^2$, respectively.

 Once the explicit expressions of the QCD spectral densities $\rho_{L/H,\pm}(s)$ are obtained,  we  take the
quark-hadron duality bellow the continuum thresholds $s_0$ and perform Borel transform  with respect to
the variable $P^2=-p^2$ to obtain  the following four QCD sum rules:
\begin{eqnarray}
\lambda_{X_{L,\pm}}^2\, \exp\left(-\frac{M_{X_{L,\pm}}^2}{T^2}\right)= \int_{4m_c^2}^{s_0} ds\, \rho_{L,\pm}(s) \, \exp\left(-\frac{s}{T^2}\right) \, , \\
\lambda_{X_{H,\mp}}^2\, \exp\left(-\frac{M_{X_{H,\mp}}^2}{T^2}\right)= \int_{4m_c^2}^{s_0} ds\, \rho_{H,\pm}(s) \, \exp\left(-\frac{s}{T^2}\right) \, ,
\end{eqnarray}
where
\begin{eqnarray}
\rho_{L,t}(s)&=&\rho_{0}(s)+\rho_{3}(s) +\rho_{4}(s)+\rho_{5}(s)+\rho_{6}(s)+\rho_{7}(s) +\rho_{8}(s)+\rho_{10}(s)\, , \\
\rho_{H,t}(s)&=&\rho_{L,t}(s)\mid_{m_c\to -m_c,t\to -t}\, ,
\end{eqnarray}

\begin{eqnarray}
\rho_{0}(s)&=&\frac{1}{3072\pi^6}\int_{y_i}^{y_f}dy \int_{z_i}^{1-y}dz \, yz(1-y-z)^3\left(s-\overline{m}_c^2\right)^2\left(35s^2-26s\overline{m}_c^2+3\overline{m}_c^4 \right) \nonumber \\
&&-\frac{3m_s m_c}{512\pi^6}\int_{y_i}^{y_f}dy \int_{z_i}^{1-y}dz \, (y+z)(1-y-z)^2\left(s-\overline{m}_c^2\right)^2\left(3s-\overline{m}_c^2 \right) \, ,
\end{eqnarray}

\begin{eqnarray}
\rho_{3}(s)&=&-\frac{m_c\langle \bar{s}s\rangle}{64\pi^4}\int_{y_i}^{y_f}dy \int_{z_i}^{1-y}dz \, (y+z)(1-y-z)\left(s-\overline{m}_c^2\right)\left(7s-3\overline{m}_c^2 \right)   \nonumber \\
&&-\frac{m_s\langle \bar{s}s\rangle}{32\pi^4}\int_{y_i}^{y_f}dy \int_{z_i}^{1-y}dz \, yz\,(1-y-z)\left(15s^2-16s\overline{m}_c^2+3\overline{m}_c^4 \right)  \nonumber \\
&&+\frac{m_s m_c^2\langle \bar{s}s\rangle}{8\pi^4}\int_{y_i}^{y_f}dy \int_{z_i}^{1-y}dz  \left(s-\overline{m}_c^2\right)\, ,
\end{eqnarray}

\begin{eqnarray}
\rho_{4}(s)&=&-\frac{m_c^2}{2304\pi^4} \langle\frac{\alpha_s GG}{\pi}\rangle\int_{y_i}^{y_f}dy \int_{z_i}^{1-y}dz \left( \frac{z}{y^2}+\frac{y}{z^2}\right)(1-y-z)^3 \left\{ 8s-3\overline{m}_c^2+s^2\,\delta\left(s-\overline{m}_c^2\right)\right\} \nonumber\\
&&+\frac{1}{1536\pi^4}\langle\frac{\alpha_s GG}{\pi}\rangle\int_{y_i}^{y_f}dy \int_{z_i}^{1-y}dz (y+z)(1-y-z)^2 \,s\,(5s-4\overline{m}_c^2) \nonumber\\
&&-\frac{tm_c^2}{1152\pi^4}\langle\frac{\alpha_s GG}{\pi}\rangle\int_{y_i}^{y_f}dy \int_{z_i}^{1-y}dz \left(s-\overline{m}_c^2 \right)\left\{ 1-\left( \frac{1}{y}+ \frac{1}{z}\right) (1-y-z) \right. \nonumber\\
&&\left.+ \frac{(1-y-z)^2}{2yz}  -\frac{1-y-z}{2} +\left(\frac{1}{y}+\frac{1}{z} \right)\frac{(1-y-z)^2}{4}
 -\frac{(1-y-z)^3}{12yz}  \right\} \nonumber \\
 &&+\frac{m_s m_c^3}{512\pi^4} \langle\frac{\alpha_s GG}{\pi}\rangle\int_{y_i}^{y_f}dy \int_{z_i}^{1-y}dz \left( \frac{1}{z^3}+\frac{1}{y^3}\right)(y+z)(1-y-z)^2 \left\{ 1+\frac{2}{3}\,s\,\delta\left(s-\overline{m}_c^2\right)\right\} \nonumber\\
 &&-\frac{m_s m_c}{512\pi^4} \langle\frac{\alpha_s GG}{\pi}\rangle\int_{y_i}^{y_f}dy \int_{z_i}^{1-y}dz \left( \frac{y}{z^2}+\frac{z}{y^2}\right)(1-y-z)^2  \left(5s-3\overline{m}_c^2\right) \nonumber\\
 &&-\frac{m_s m_c}{768\pi^4} \langle\frac{\alpha_s GG}{\pi}\rangle\int_{y_i}^{y_f}dy \int_{z_i}^{1-y}dz \, (1-y-z)  \left(5s-3\overline{m}_c^2\right) \nonumber\\
 &&-\frac{tm_s m_c}{1152\pi^4} \langle\frac{\alpha_s GG}{\pi}\rangle\int_{y_i}^{y_f}dy \int_{z_i}^{1-y}dz \, (1-y-z)  \left(5s-3\overline{m}_c^2\right) \nonumber\\
 &&+\frac{tm_s m_c}{4608\pi^4} \langle\frac{\alpha_s GG}{\pi}\rangle\int_{y_i}^{y_f}dy \int_{z_i}^{1-y}dz \left(\frac{1}{y}+\frac{1}{z} \right) (1-y-z)^2  \left(5s-3\overline{m}_c^2\right) \, ,
\end{eqnarray}

\begin{eqnarray}
\rho_{5}(s)&=&\frac{m_c\langle \bar{s}g_s\sigma Gs\rangle}{128\pi^4}\int_{y_i}^{y_f}dy \int_{z_i}^{1-y}dz   (y+z) \left(5s-3\overline{m}_c^2 \right) \nonumber\\
&&-\frac{m_c\langle \bar{s}g_s\sigma Gs\rangle}{128\pi^4}\int_{y_i}^{y_f}dy \int_{z_i}^{1-y}dz   \left(\frac{y}{z}+\frac{z}{y} \right)(1-y-z) \left(2s-\overline{m}_c^2 \right)  \nonumber\\
&&-\frac{tm_c\langle \bar{s}g_s\sigma Gs\rangle}{1152\pi^4}\int_{y_i}^{y_f}dy \int_{z_i}^{1-y}dz   \left(\frac{y}{z}+\frac{z}{y} \right)(1-y-z) \left(5s-3\overline{m}_c^2 \right)   \nonumber\\
&&-\frac{m_s m_c^2\langle \bar{s}g_s\sigma Gs\rangle}{32\pi^4}\int_{y_i}^{y_f}dy   \nonumber\\
&&+ \frac{m_s\langle\bar{s}g_s\sigma Gs\rangle }{96\pi^4}\int_{y_i}^{y_f}dy \int_{z_i}^{1-y}dz\, yz \left\{8s-3\overline{m}_c^2 +s^2\,\delta\left(s-\overline{m}_c^2 \right)\right\}\nonumber\\
&&+\frac{m_s m^2_c\langle \bar{s}g_s\sigma Gs\rangle}{128\pi^4}\int_{y_i}^{y_f}dy \int_{z_i}^{1-y}dz   \left(\frac{1}{y}+\frac{1}{z} \right)  \nonumber\\
&&+\frac{t m_s\langle \bar{s}g_s\sigma Gs\rangle}{1152\pi^4}\int_{y_i}^{y_f}dy \int_{z_i}^{1-y}dz   \,(y+z) \left(5s-3\overline{m}_c^2 \right) \, ,
\end{eqnarray}

\begin{eqnarray}
\rho_{6}(s)&=&\frac{m_c^2\langle\bar{s}s\rangle^2}{12\pi^2}\int_{y_i}^{y_f}dy +\frac{g_s^2\langle\bar{s}s\rangle^2}{648\pi^4}\int_{y_i}^{y_f}dy \int_{z_i}^{1-y}dz\, yz \left\{8s-3\overline{m}_c^2 +s^2\,\delta\left(s-\overline{m}_c^2 \right)\right\}\nonumber\\
&&-\frac{g_s^2\langle\bar{s}s\rangle^2}{2592\pi^4}\int_{y_i}^{y_f}dy \int_{z_i}^{1-y}dz(1-y-z)\left\{ \left(\frac{z}{y}+\frac{y}{z} \right)3\left(7s-4\overline{m}_c^2 \right)\right.\nonumber\\
&&\left.+\left(\frac{z}{y^2}+\frac{y}{z^2} \right)m_c^2\left[ 7+5s\,\delta\left(s-\overline{m}_c^2 \right)\right]-(y+z)\left(4s-3\overline{m}_c^2 \right)\right\} \nonumber\\
&&-\frac{g_s^2\langle\bar{s}s\rangle^2}{3888\pi^4}\int_{y_i}^{y_f}dy \int_{z_i}^{1-y}dz(1-y-z)\left\{  \left(\frac{z}{y}+\frac{y}{z} \right)3\left(2s-\overline{m}_c^2 \right)\right. \nonumber\\
&&\left.+\left(\frac{z}{y^2}+\frac{y}{z^2} \right)m_c^2\left[ 1+s\,\delta\left(s-\overline{m}_c^2\right)\right]+(y+z)2\left[8s-3\overline{m}_c^2 +s^2\,\delta\left(s-\overline{m}_c^2\right)\right]\right\} \nonumber\\
&&-\frac{m_s m_c g_s^2\langle\bar{s}s\rangle^2}{864\pi^4}\int_{y_i}^{y_f}dy \left\{1 +\frac{2}{3}\,s\,\delta\left(s-\widetilde{m}_c^2 \right)\right\}\nonumber\\
&&+\frac{m_s m_c g_s^2\langle\bar{s}s\rangle^2}{2592\pi^4}\int_{y_i}^{y_f}dy \int_{z_i}^{1-y}dz\left\{\frac{9-y}{y}+\frac{9-z}{z} +5m_c^2\left(\frac{1}{y^2}+\frac{1}{z^2}\right)\delta\left(s-\overline{m}_c^2 \right)\right\}\nonumber\\
&&+\frac{m_s m_c g_s^2\langle\bar{s}s\rangle^2}{864\pi^4}\int_{y_i}^{y_f}dy\int_{z_i}^{1-y}dz \,\left( \frac{y}{z}+\frac{z}{y}\right)\left\{1 +\frac{2}{3}\,s\,\delta\left(s-\overline{m}_c^2 \right)\right\} \nonumber\\
&&+\frac{m_s m_c  \langle\bar{s}s\rangle^2}{16\pi^2}\int_{y_i}^{y_f}dy  \, \left\{1 +\frac{2}{3}\,s\,\delta\left(s-\widetilde{m}_c^2 \right)\right\} \, ,
\end{eqnarray}

\begin{eqnarray}
\rho_7(s)&=&\frac{m_c^3\langle\bar{s}s\rangle}{576\pi^2}\langle\frac{\alpha_sGG}{\pi}\rangle\int_{y_i}^{y_f}dy \int_{z_i}^{1-y}dz \left(\frac{1}{z^3}+\frac{1}{y^3} \right)(y+z)(1-y-z) \left( 1+\frac{2s}{T^2}\right)\delta\left(s-\overline{m}_c^2\right)\nonumber\\
&&-\frac{m_c\langle\bar{s}s\rangle}{64\pi^2}\langle\frac{\alpha_sGG}{\pi}\rangle\int_{y_i}^{y_f}dy \int_{z_i}^{1-y}dz \left(\frac{y}{z^2}+\frac{z}{y^2}\right)(1-y-z)
\left\{1+\frac{2s}{3}\delta\left(s-\overline{m}_c^2\right) \right\} \nonumber\\
&&-\frac{m_c\langle\bar{s}s\rangle}{192\pi^2}\langle\frac{\alpha_sGG}{\pi}\rangle\int_{y_i}^{y_f}dy \int_{z_i}^{1-y}dz\left\{1+\frac{2s}{3}\delta\left(s-\overline{m}_c^2\right) \right\} \nonumber\\
&&-\frac{tm_c\langle\bar{s}s\rangle}{288\pi^2}\langle\frac{\alpha_sGG}{\pi}\rangle\int_{y_i}^{y_f}dy \int_{z_i}^{1-y}dz\left\{1-\left(\frac{1}{y}+\frac{1}{z}\right)\frac{1-y-z}{2}\right\}\left\{1+\frac{2s}{3}\delta\left(s-\overline{m}_c^2\right) \right\} \nonumber\\
&&-\frac{m_c\langle\bar{s}s\rangle}{384\pi^2}\langle\frac{\alpha_sGG}{\pi}\rangle\int_{y_i}^{y_f}dy \left\{1+\frac{2s}{3}\delta\left(s-\widetilde{m}_c^2\right) \right\}    \nonumber\\
&&+\frac{m_s m_c^2 \langle\bar{s}s\rangle}{288\pi^2T^2}\langle\frac{\alpha_sGG}{\pi}\rangle\int_{y_i}^{y_f}dy \int_{z_i}^{1-y}dz \left(\frac{y}{z^2}+\frac{z}{y^2}\right)(1-y-z)
\left(s+\frac{s^2}{T^2} \right) \delta\left(s-\overline{m}_c^2\right)\nonumber\\
&&-\frac{m_s m_c^4 \langle\bar{s}s\rangle}{144\pi^2T^2}\langle\frac{\alpha_sGG}{\pi}\rangle\int_{y_i}^{y_f}dy \int_{z_i}^{1-y}dz \left(\frac{1}{z^3}+\frac{1}{y^3}\right)  \delta\left(s-\overline{m}_c^2\right)\nonumber\\
&&+\frac{m_s m_c^2 \langle\bar{s}s\rangle}{48\pi^2}\langle\frac{\alpha_sGG}{\pi}\rangle\int_{y_i}^{y_f}dy \int_{z_i}^{1-y}dz \left(\frac{1}{z^2}+\frac{1}{y^2}\right)  \delta\left(s-\overline{m}_c^2\right)\nonumber\\
&&-\frac{m_s   \langle\bar{s}s\rangle}{576\pi^2}\langle\frac{\alpha_sGG}{\pi}\rangle\int_{y_i}^{y_f}dy \int_{z_i}^{1-y}dz \,\left(y+z \right) \left(1+\frac{s}{2T^2}\right)  \delta\left(s-\overline{m}_c^2\right)\nonumber\\
&&+\frac{tm_s m_c^2   \langle\bar{s}s\rangle}{3456\pi^2}\langle\frac{\alpha_sGG}{\pi}\rangle\int_{y_i}^{y_f}dy \int_{z_i}^{1-y}dz \,\frac{2+3y+3z}{yz}   \delta\left(s-\overline{m}_c^2\right)\nonumber\\
&&-\frac{tm_s m_c^2   \langle\bar{s}s\rangle}{1728\pi^2}\langle\frac{\alpha_sGG}{\pi}\rangle\int_{y_i}^{y_f}dy \,\left(\frac{1}{y}+\frac{1}{1-y} \right)   \delta\left(s-\widetilde{m}_c^2\right)\nonumber\\
&&-\frac{tm_s    \langle\bar{s}s\rangle}{288\pi^2}\langle\frac{\alpha_sGG}{\pi}\rangle\int_{y_i}^{y_f}dy \int_{z_i}^{1-y}dz \,\left\{1+\frac{2}{3}s\,\delta\left(s-\overline{m}_c^2\right) \right\}  \, ,
\end{eqnarray}

\begin{eqnarray}
\rho_8(s)&=&-\frac{m_c^2\langle\bar{s}s\rangle\langle\bar{s}g_s\sigma Gs\rangle}{24\pi^2}\int_0^1 dy \left(1+\frac{s}{T^2} \right)\delta\left(s-\widetilde{m}_c^2\right)\nonumber\\
&&+\frac{m_c^2\langle\bar{s}s\rangle\langle\bar{s}g_s\sigma Gs\rangle}{96\pi^2}\int_0^1 dy \left( \frac{1}{y}+\frac{1}{1-y} \right)\delta\left(s-\widetilde{m}_c^2\right)\nonumber\\
&&+\frac{t\langle\bar{s}s\rangle\langle\bar{s}g_s\sigma Gs\rangle}{288\pi^2}\int_{y_i}^{y_f} dy \left\{1+\frac{2s}{3}\delta\left(s-\widetilde{m}_c^2\right) \right\}   \nonumber \\
&&-\frac{5m_s m_c\langle\bar{s}s\rangle\langle\bar{s}g_s\sigma Gs\rangle}{288\pi^2}\int_{y_i}^{y_f} dy \left(1+\frac{3s}{2T^2}+\frac{s^2}{T^4} \right) \delta\left(s-\widetilde{m}_c^2\right)   \nonumber \\
&&+\frac{m_s m_c\langle\bar{s}s\rangle\langle\bar{s}g_s\sigma Gs\rangle}{192\pi^2 T^2}\int_{y_i}^{y_f} dy \left(\frac{1-y}{y}+\frac{y}{1-y} \right) \,s\,\delta\left(s-\widetilde{m}_c^2\right)   \nonumber \\
&&+\frac{tm_s m_c\langle\bar{s}s\rangle\langle\bar{s}g_s\sigma Gs\rangle}{1728\pi^2 }\int_{y_i}^{y_f} dy \left(\frac{1-y}{y}+\frac{y}{1-y} \right) \,\left( 1+ \frac{2s}{T^2}\right)\,\delta\left(s-\widetilde{m}_c^2\right)   \, ,
\end{eqnarray}

\begin{eqnarray}
\rho_{10}(s)&=&\frac{m_c^2\langle\bar{s}g_s\sigma Gs\rangle^2}{192\pi^2T^6}\int_0^1 dy \, s^2\,\delta \left( s-\widetilde{m}_c^2\right)\nonumber\\
&&-\frac{m_c^4\langle\bar{s}s\rangle^2}{216T^4}\langle\frac{\alpha_sGG}{\pi}\rangle\int_0^1 dy  \left\{ \frac{1}{y^3}+\frac{1}{(1-y)^3}\right\} \delta\left( s-\widetilde{m}_c^2\right)\nonumber\\
&&+\frac{m_c^2\langle\bar{s}s\rangle^2}{72T^2}\langle\frac{\alpha_sGG}{\pi}\rangle\int_0^1 dy  \left\{ \frac{1}{y^2}+\frac{1}{(1-y)^2}\right\} \delta\left( s-\widetilde{m}_c^2\right)\nonumber\\
&&-\frac{t\langle\bar{s}s\rangle^2}{1296}\langle\frac{\alpha_sGG}{\pi}\rangle\int_0^1 dy  \left( 1+\frac{2s}{T^2}\right) \delta\left( s-\widetilde{m}_c^2\right)\nonumber\\
&&-\frac{m_c^2\langle\bar{s}g_s\sigma Gs\rangle^2}{384\pi^2T^4}\int_0^1 dy \left( \frac{1}{y}+\frac{1}{1-y}\right)\,s\,\delta \left( s-\widetilde{m}_c^2\right)\nonumber\\
&&-\frac{t\langle\bar{s}g_s\sigma Gs\rangle^2}{1728\pi^2}\int_0^1 dy \left(1+\frac{3s}{2T^2}+\frac{s^2}{T^4} \right)\delta \left( s-\widetilde{m}_c^2\right)\nonumber\\
&&-\frac{t\langle\bar{s}g_s\sigma Gs\rangle^2}{2304\pi^2}\int_0^1 dy \left(1+\frac{2s}{T^2}  \right)\delta \left( s-\widetilde{m}_c^2\right)\nonumber\\
&&+\frac{m_c^2\langle\bar{s}s\rangle^2}{216T^6}\langle\frac{\alpha_sGG}{\pi}\rangle\int_0^1 dy  \,s^2\,  \delta\left( s-\widetilde{m}_c^2\right) \nonumber\\
&&-\frac{m_s m_c\langle\bar{s}g_s\sigma Gs\rangle^2}{576\pi^2 T^2}\int_0^1 dy \left(1+\frac{s}{T^2}+\frac{s^2}{2T^4}-\frac{s^3}{T^6}  \right)\delta \left( s-\widetilde{m}_c^2\right)\nonumber\\
&&+\frac{m_s m_c^3\langle\bar{s}s\rangle^2}{288T^4}\langle\frac{\alpha_sGG}{\pi}\rangle\int_0^1 dy \left[ \frac{1}{(1-y)^3}+\frac{1}{y^3}\right]  \left( 1-\frac{2s}{3T^2}\right) \delta\left( s-\widetilde{m}_c^2\right)\nonumber\\
&&-\frac{m_s m_c\langle\bar{s}s\rangle^2}{288T^2}\langle\frac{\alpha_sGG}{\pi}\rangle\int_0^1 dy \left[ \frac{y}{(1-y)^2}+\frac{1-y}{y^2}\right]  \left( 1-\frac{2s}{T^2}\right) \delta\left( s-\widetilde{m}_c^2\right)\nonumber\\
&&+\frac{t m_s m_c\langle\bar{s}s\rangle^2}{2592T^2}\langle\frac{\alpha_sGG}{\pi}\rangle\int_0^1 dy \left( \frac{1}{y}+\frac{1}{1-y}\right)  \left( 1-\frac{2s}{T^2}\right) \delta\left( s-\widetilde{m}_c^2\right)\nonumber\\
&&+\frac{m_s m_c \langle\bar{s}g_s\sigma Gs\rangle^2}{1152\pi^2T^2}  \int_0^1 dy \left( \frac{1-y}{y}+\frac{y}{1-y}\right)  \left( 1+\frac{s}{T^2}-\frac{s^2}{T^4}\right) \delta\left( s-\widetilde{m}_c^2\right)\nonumber\\
&&+\frac{t m_s m_c \langle\bar{s}g_s\sigma Gs\rangle^2}{10368\pi^2T^2}  \int_0^1 dy \left( \frac{1-y}{y}+\frac{y}{1-y}\right)  \left( 1+\frac{s}{T^2}-\frac{2s^2}{T^4}\right) \delta\left( s-\widetilde{m}_c^2\right)\nonumber\\
&&-\frac{ m_s m_c\langle\bar{s}s\rangle^2}{864T^2}\langle\frac{\alpha_sGG}{\pi}\rangle\int_0^1 dy    \left( 1+\frac{s}{T^2}+\frac{s^2}{2T^4}-\frac{s^3}{T^6}\right) \delta\left( s-\widetilde{m}_c^2\right) \, ,
\end{eqnarray}
where the $T^2$ is the Borel parameter,  $y_{f}=\frac{1+\sqrt{1-4m_c^2/s}}{2}$,
$y_{i}=\frac{1-\sqrt{1-4m_c^2/s}}{2}$, $z_{i}=\frac{ym_c^2}{y s -m_c^2}$, $\overline{m}_c^2=\frac{(y+z)m_c^2}{yz}$,
$ \widetilde{m}_c^2=\frac{m_c^2}{y(1-y)}$, $\int_{y_i}^{y_f}dy \to \int_{0}^{1}dy$, $\int_{z_i}^{1-y}dz \to \int_{0}^{1-y}dz$ when the $\delta$ functions $\delta\left(s-\overline{m}_c^2\right)$ and $\delta\left(s-\widetilde{m}_c^2\right)$ appear.

 In this article, we carry out the
operator product expansion for the vacuum condensates  up to dimension 10, and
  assume  vacuum saturation for the  higher dimension vacuum condensates. The vacuum condensates  are   the vacuum expectations
of the operators,  we take the truncations $n\leq 10$ and $k\leq 1$ for the operators in a consistent way,
and discard the operators of the orders $\mathcal{O}( \alpha_s^{k})$ with $k> 1$.
The terms of the orders $\mathcal{O}(\frac{1}{T^2})$, $\mathcal{O}(\frac{1}{T^4})$, $\mathcal{O}(\frac{1}{T^6})$, $\mathcal{O}(\frac{1}{T^8})$ in the QCD spectral densities manifest themselves at small  $T^2$, we have to choose large  $T^2$ to warrant convergence of the operator product expansion and appearance of the Borel platforms. The higher dimension vacuum condensates play an important role in determining the Borel windows, though they play a less important role in the Borel windows.

 We differentiate   Eqs.(17-18) with respect to  $\frac{1}{T^2}$, then eliminate the
 pole residues $\lambda_{X_{L,\pm}}$ and $\lambda_{X_{H,\pm}}$, and obtain the QCD sum rules for
 the masses of the $X_{L,\pm}$ and $X_{H,\pm}$, respectively.
 \begin{eqnarray}
 M^2_{X_{L,\pm}}&=& -\frac{\int_{4m_c^2}^{s_0} ds\frac{d}{d \left(1/T^2\right)}\,\rho_{L,\pm}(s)\exp\left(-\frac{s}{T^2}\right)}{\int_{4m_c^2}^{s_0} ds \,\rho_{L,\pm}(s)
 \exp\left(-\frac{s}{T^2}\right)}\,  , \\
 M^2_{X_{H,\mp}}&=& -\frac{\int_{4m_c^2}^{s_0} ds\frac{d}{d \left(1/T^2\right)}\,\rho_{H,\pm}(s)\exp\left(-\frac{s}{T^2}\right)}{\int_{4m_c^2}^{s_0} ds \,\rho_{H,\pm}(s)
 \exp\left(-\frac{s}{T^2}\right)}\,  .
\end{eqnarray}

\section{Numerical results and discussions}

In previous works, we described   the hidden-charm and the hidden-bottom four-quark systems  $q\bar{q}^{\prime}Q\bar{Q}$
by a double-well potential \cite{WangHuangTao-3900,WangTetraquarkCTP,Wang-4660-2014,Wang-Huang-NPA-2014}.     In the four-quark system $q\bar{q}^{\prime}Q\bar{Q}$,
 the heavy quark $Q$ serves as one static well potential and  combines with the light quark $q$  to form a heavy diquark $\mathcal{D}_{qQ}$ in  color antitriplet,
while the heavy antiquark $\bar{Q}$ serves  as the other static well potential and combines with the light antiquark $\bar{q}^\prime$  to form a heavy antidiquark $\mathcal{D}_{\bar{q}^{\prime}\bar{Q}}$ in  color triplet.
 Then  the  $\mathcal{D}_{qQ}$ and $\mathcal{D}_{\bar{q}^{\prime}\bar{Q}}$ combine together to form a compact tetraquark state,
the two heavy quarks $Q$ and $\bar{Q}$ stabilize the tetraquark state \cite{Brodsky-2014}.

The doubly-heavy tetraquark states  are characterized by the effective heavy quark mass ${\mathbb{M}}_Q$ and the virtuality $V=\sqrt{M^2_{X/Y/Z}-(2{\mathbb{M}}_Q)^2}$. It is natural to take the energy  scale $\mu=V$,
 the energy scale formula works well for the  $X(3872)$, $Z_c(3900)$,  $Z_c(4020)$, $Z_c(4025)$, $Z(4430)$, $Y(4660)$, $Z_b(10610)$  and $Z_b(10650)$ in the  scenario of  tetraquark  states \cite{WangHuangTao-3900,WangTetraquarkCTP,Wang-4660-2014,Wang-Huang-NPA-2014,Wang-3900-4430,Wang-1601}. In Refs.\cite{WangHuangTao-3900,Wang-4660-2014}, we obtain the  effective mass  for  the diquark-antidiquark type hidden-charm tetraquark states,   ${\mathbb{M}}_c=1.8\,\rm{GeV}$. Then we re-checked the numerical calculations  and found that there exists  a small error involving the mixed condensates.  After the small error is corrected, the Borel windows are modified slightly and the numerical results are  improved slightly,  the conclusions survive. In this article, we choose the updated value ${\mathbb{M}}_c=1.82\,\rm{GeV}$ \cite{Wang-1601}, and obtain the optimal energy scales $\mu=1.4\,\rm{GeV}$ and $2.0\,\rm{GeV}$ for the QCD spectral densities of the QCD sum rules for the  $Z_c(3900)$ and $X(4140)$, respectively.

Now we choose the input parameters at the QCD side of the QCD sum rules.
We take the vacuum condensates  to be the standard values
$\langle\bar{q}q \rangle=-(0.24\pm 0.01\, \rm{GeV})^3$,  $\langle\bar{s}s \rangle=(0.8\pm0.1)\langle\bar{q}q \rangle$,
$\langle\bar{q}g_s\sigma G q \rangle=m_0^2\langle \bar{q}q \rangle$, $\langle\bar{s}g_s\sigma G s \rangle=m_0^2\langle \bar{s}s \rangle$,
$m_0^2=(0.8 \pm 0.1)\,\rm{GeV}^2$, $\langle \frac{\alpha_s
GG}{\pi}\rangle=(0.33\,\rm{GeV})^4 $    at the energy scale  $\mu=1\, \rm{GeV}$
\cite{SVZ79,Reinders85,ColangeloReview}, and  take the $\overline{MS}$ masses $m_{c}(m_c)=(1.275\pm0.025)\,\rm{GeV}$ and $m_s(\mu=2\,\rm{GeV})=(0.095\pm0.005)\,\rm{GeV}$
 from the Particle Data Group \cite{PDG}.
Moreover,  we take into account
the energy-scale dependence of  the quark condensates, mixed quark condensates and $\overline{MS}$ masses from the renormalization group equation \cite{Narison-book},
 \begin{eqnarray}
\langle\bar{q}q \rangle(\mu)&=&\langle\bar{q}q \rangle(Q)\left[\frac{\alpha_{s}(Q)}{\alpha_{s}(\mu)}\right]^{\frac{4}{9}}\, , \nonumber\\
 \langle\bar{s}s \rangle(\mu)&=&\langle\bar{s}s \rangle(Q)\left[\frac{\alpha_{s}(Q)}{\alpha_{s}(\mu)}\right]^{\frac{4}{9}}\, , \nonumber\\
 \langle\bar{q}g_s \sigma Gq \rangle(\mu)&=&\langle\bar{q}g_s \sigma Gq \rangle(Q)\left[\frac{\alpha_{s}(Q)}{\alpha_{s}(\mu)}\right]^{\frac{2}{27}}\, , \nonumber\\ \langle\bar{s}g_s \sigma Gs \rangle(\mu)&=&\langle\bar{s}g_s \sigma Gs \rangle(Q)\left[\frac{\alpha_{s}(Q)}{\alpha_{s}(\mu)}\right]^{\frac{2}{27}}\, ,
\end{eqnarray}
\begin{eqnarray}
m_c(\mu)&=&m_c(m_c)\left[\frac{\alpha_{s}(\mu)}{\alpha_{s}(m_c)}\right]^{\frac{12}{25}} \, ,\nonumber\\
m_s(\mu)&=&m_s({\rm 2GeV} )\left[\frac{\alpha_{s}(\mu)}{\alpha_{s}({\rm 2GeV})}\right]^{\frac{4}{9}} \, ,\nonumber\\
\alpha_s(\mu)&=&\frac{1}{b_0t}\left[1-\frac{b_1}{b_0^2}\frac{\log t}{t} +\frac{b_1^2(\log^2{t}-\log{t}-1)+b_0b_2}{b_0^4t^2}\right]\, ,
\end{eqnarray}
  where $t=\log \frac{\mu^2}{\Lambda^2}$, $b_0=\frac{33-2n_f}{12\pi}$, $b_1=\frac{153-19n_f}{24\pi^2}$, $b_2=\frac{2857-\frac{5033}{9}n_f+\frac{325}{27}n_f^2}{128\pi^3}$,  $\Lambda=213\,\rm{MeV}$, $296\,\rm{MeV}$  and  $339\,\rm{MeV}$ for the flavors  $n_f=5$, $4$ and $3$, respectively  \cite{PDG}. In this article, we take the standard  value of the quark condensate $\langle\bar{q}q\rangle$ at the energy scale $\mu=1\,\rm{GeV}$ from the Gell-Mann-Oakes-Renner relation \cite{SVZ79,Reinders85,ColangeloReview,Narison-book,GMOR}. The values of the quark condensates have been updated \cite{WangYM},  however, we determine the effective heavy quark masses ${\mathbb{M}}_Q$ with the standard values \cite{WangHuangTao-3900,WangTetraquarkCTP,Wang-4660-2014,Wang-Huang-NPA-2014,Wang-3900-4430,Wang-1601}, so we choose  the standard values in this article. In our next works, we will redetermine the ${\mathbb{M}}_Q$ with the updated values, as the updated value $\langle\bar{q}q\rangle({2\rm GeV })=-(274^{+15}_{-17}\,{\rm MeV})^3$ differs from the standard value $\langle\bar{q}q\rangle({2\rm GeV })=-(257 \pm 10\,{\rm MeV})^3$ considerably.

  In this article, we have neglected the higher-order QCD corrections. Including the  higher-order QCD corrections means refitting   the  effective $c$-quark mass ${\mathbb{M}}_c$. According to the energy scale formula $\mu=\sqrt{M^2_{X/Y/Z}-(2{\mathbb{M}}_c)^2}$, some uncertainties are introduced by neglecting the higher-order QCD corrections.  In this article, we take the leading order approximations just as in the QCD sum rules for the $X(3872)$, $Z_c(3900)$, $Y(4660)$, some higher-order effects are embodied in the effective $c$-quark mass ${\mathbb{M}}_c$ \cite{WangHuangTao-3900,Wang-4660-2014}.

 In Ref.\cite{Wang-3900-4430}, we  observed that  the $Z_c(3900)$ and $Z(4430)$ can be assigned to be the ground state and the first radial excited state of the axialvector  tetraquark states with $J^{PC}=1^{+-}$, respectively based on the QCD sum rules. We expect the energy gap between the ground state and the first radial excited state of the hidden-charm  tetraquark states is about $0.6\,\rm{GeV}$ according to the mass difference $M_{Z(4430)}-M_{Z_c(3900)}=576\,\rm{MeV}$.  In this article, we assume $X(4140)=X_{L,+}$, then the threshold parameters can be taken as $\sqrt{s_0}=(4.6-4.8)\,\rm{GeV}$. If we choose the  energy scale determined by the empirical energy scale formula, then $\mu=2.0\,\rm{GeV}$. In calculations, we observe that it is impossible to reproduce the experimental value $M_{X(4140)} = 4146.5 \pm 4.5 ^{+4.6}_{-2.8} \mbox{ MeV}$.

 Now we explore the energy scale dependence of the predicted mass of the $X_{L,+}$. In Fig.1,  we plot the mass  with variation of the  Borel parameter $T^2$ and energy scale $\mu$ for the
threshold parameter $\sqrt{s_0}=4.7\,\rm{GeV}$. From the figure, we can see that the masses decrease monotonously
with increase of the energy scales. The energy scale $\mu=1.1\,\rm{GeV}$ is the optimal energy scale to reproduce the experimental value.
If we choose the energy scale $\mu=1.1\,\rm{GeV}$ and threshold parameter $\sqrt{s_0}=(4.6-4.8)\,\rm{GeV}$, the ideal Borel parameter is $T^2=(2.5-2.9)\,\rm{GeV}^2$, the pole contribution is about $(52-75)\%$, the contributions of the vacuum condensates of dimension 8 and 10 are about $-(9-16)\%$ and $1\ll\%$, respectively.  The  two criteria of the QCD sum rules (i.e. pole dominance at the phenomenological side and convergence of the operator product expansion at the QCD side) are both  satisfied. After taking into account all uncertainties  of the input parameters, we obtain the mass and pole residue,
\begin{eqnarray}
M_{X_{L,+}}&=&(4.15\pm 0.09) \,\rm{GeV} \, ,  \nonumber\\
\lambda_{X_{L,+}}&=&(2.10\pm0.30)\times 10^{-2}\,\rm{GeV}^5 \, ,
\end{eqnarray}
  which are shown in Fig.2 at a large interval of the Borel parameter. The predicted mass $M_{X_{L,+}}=(4.15\pm 0.09) \,\rm{GeV}$ is in excellent agreement with the experimental value $M_{X(4140)} = 4146.5 \pm 4.5 ^{+4.6}_{-2.8} \mbox{ MeV}$, which favors assigning the $X(4140)$ to be the $1^{++}$ diquark-antidiquark type $cs\bar{c}\bar{s}$  tetraquark state. However, we reproduce the experimental value $M_{Z_c(3900)}$ at the energy scale $\mu=1.4\,\rm{GeV}$ of the QCD spectral density, while we reproduce the experimental value $M_{X(4140)}$ at the energy scale $\mu=1.1\,\rm{GeV}$ of the QCD spectral density.
  The empirical energy scale formula can be re-written as
\begin{eqnarray}
M^2_{X/Y/Z}=(2{\mathbb{M}}_Q)^2+\mu^2 \, ,
\end{eqnarray}
which puts  a strong constraint on the masses of the hidden-charm and the hidden-bottom  tetraquark states.
  If the two heavy quarks $Q$ and $\bar{Q}$ serve as a double-well potential and stabilize the tetraquark states, the $X(4140)$ should correspond to a larger energy scale than that of the $Z_c(3900)$, i.e. $\mu_{X(4140)}>\mu_{Z_c(3900)}$. Moreover,  in previous works, we used the empirical energy scale formula and reproduced the experimental values of the  masses of the  $X(3872)$, $Z_c(3900)$,  $Z_c(4020)$, $Z_c(4025)$, $Z(4430)$, $Y(4660)$, $Z_b(10610)$  and $Z_b(10650)$ in the  scenario of  tetraquark  states  \cite{WangHuangTao-3900,WangTetraquarkCTP,Wang-4660-2014,Wang-Huang-NPA-2014,Wang-3900-4430,Wang-1601}. It is odd that the QCD spectral density of the QCD sum rules for the $X(4140)$ does not obey the empirical energy scale formula.

  Now  we search for the  Borel parameters $T^2$ and continuum threshold
parameters $s_0$  to satisfy the  following four  criteria:

$\bf{1_\cdot}$ Pole dominance at the phenomenological side;

$\bf{2_\cdot}$ Convergence of the operator product expansion;

$\bf{3_\cdot}$ Appearance of the Borel platforms;

$\bf{4_\cdot}$ Satisfying the energy scale formula, \\
to obtain the ground state masses of the $X_{L,\pm}$ and $X_{H,\pm}$.

The resulting Borel parameters, continuum threshold parameters, energy scales, pole contributions, contributions of the vacuum condensates of  dimension 8  and 10 are shown explicitly in Table 1, where  the vacuum condensate contributions $D_{8}$ and $D_{10}$ correspond to the central values of the threshold parameters. From the Table, we can see that the first two criteria are satisfied.

We take into account all uncertainties of the input parameters, and obtain the values of the ground state masses and pole residues, which are shown explicitly in Table 2 and Figs.3-4. From Table 1 and Table 2, we can see that the empirical energy scale formula is satisfied. From Figs.3-4, we can see that in the Borel windows, the masses and pole residues are rather stable with variations of the Borel parameters. The four criteria are all satisfied, we expect to make reliable predictions.  From Fig.3, we can see that the upper error bound of the  theoretical value $M_{X_{L,+}}$ lies below the experimental value $M_{X(4140)}$, the present prediction disfavors  assigning the $X(4140)$ to be diquark-antidiquark type  $cs\bar{c}\bar{s}$ tetraquark state with the $J^{PC}=1^{++}$. The present predictions of the masses of the axialvector $cs\bar{c}\bar{s}$ tetraquark states can be confronted to the  experimental data in the future.

\begin{figure}
\centering
\includegraphics[totalheight=7cm,width=8cm]{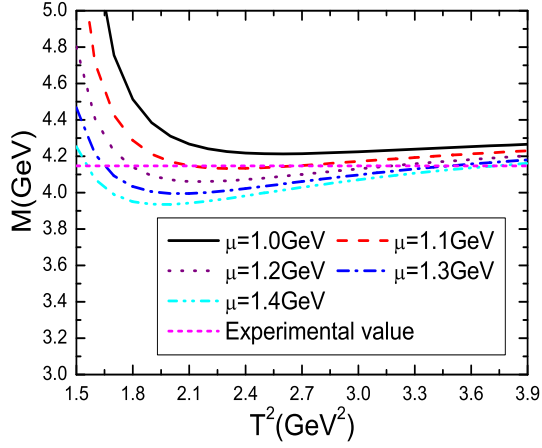}
  \caption{ The masses $M_{X_{L,+}}$ with variations of the  Borel parameters $T^2$ and energy scales $\mu$, where the horizontal line denotes the experimental value of the mass $M_{X(4140)}$. }
\end{figure}

\begin{figure}
\centering
\includegraphics[totalheight=6cm,width=7cm]{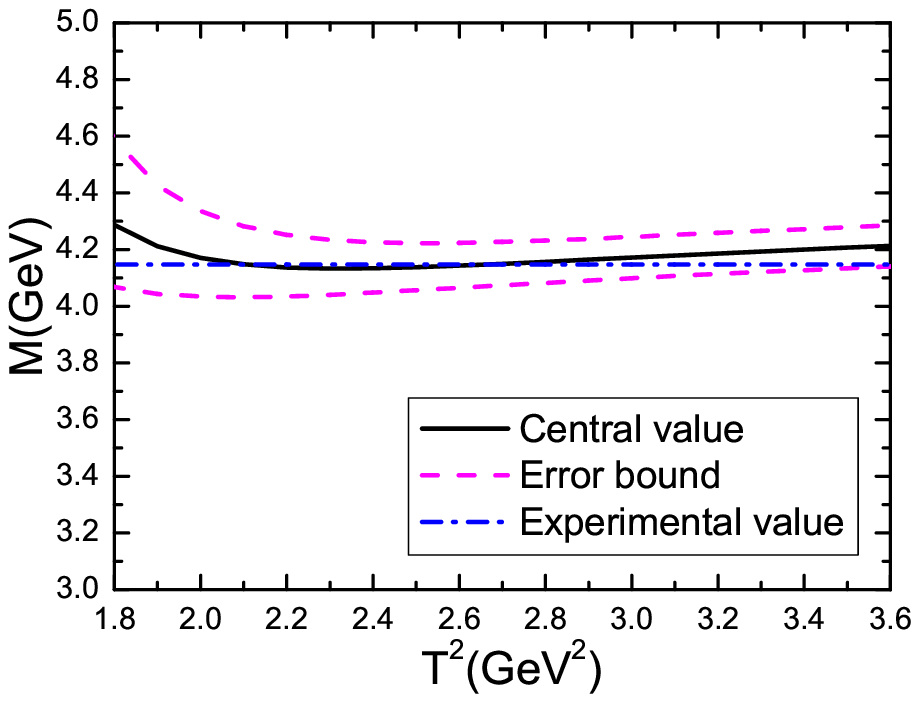}
\includegraphics[totalheight=6cm,width=7cm]{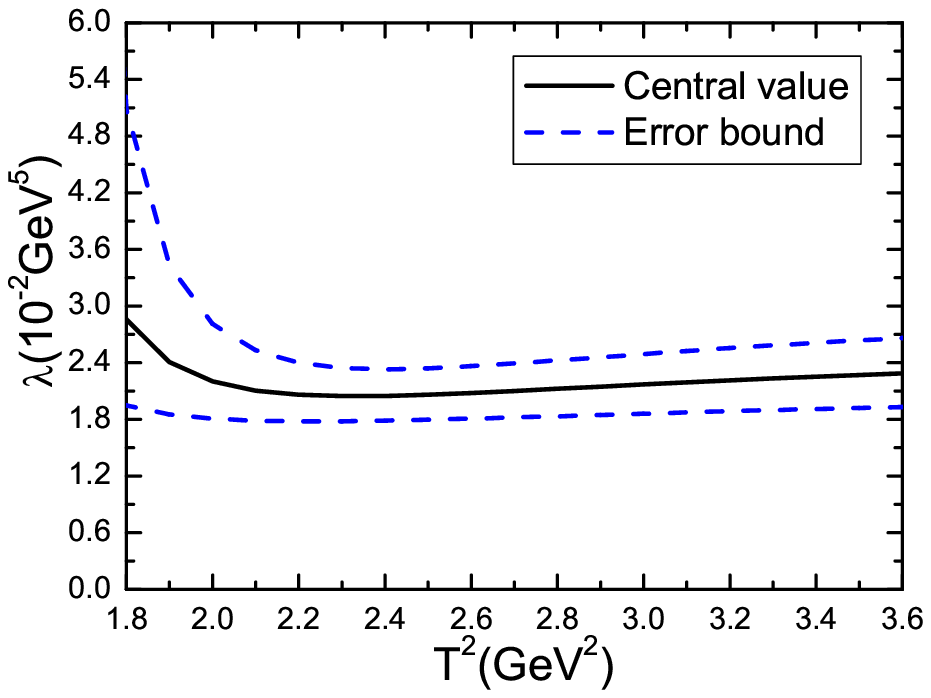}
  \caption{ The mass and pole residue of the $X_{L,+}$  with variations of the  Borel parameter $T^2$, where the horizontal line denotes the experimental value of the mass $M_{X(4140)}$. }
\end{figure}

\begin{figure}
\centering
\includegraphics[totalheight=6cm,width=7cm]{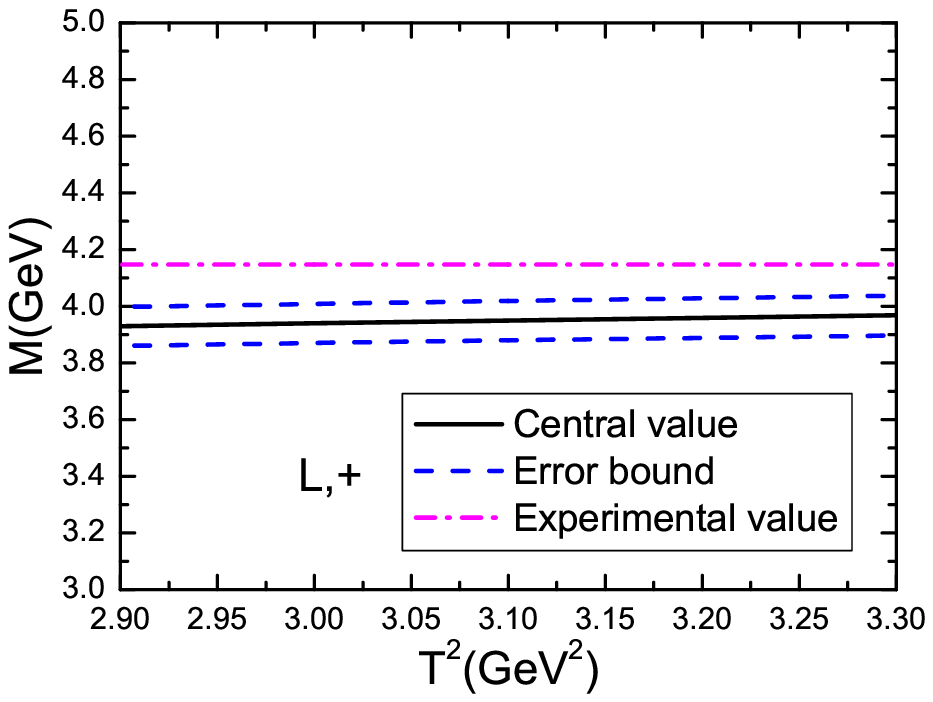}
\includegraphics[totalheight=6cm,width=7cm]{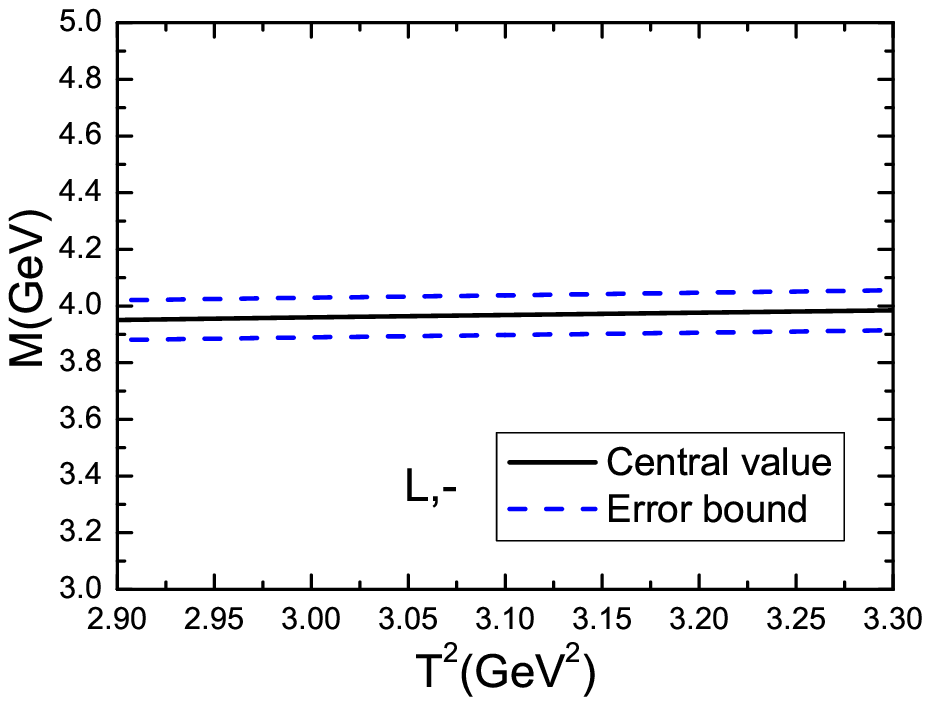}
\includegraphics[totalheight=6cm,width=7cm]{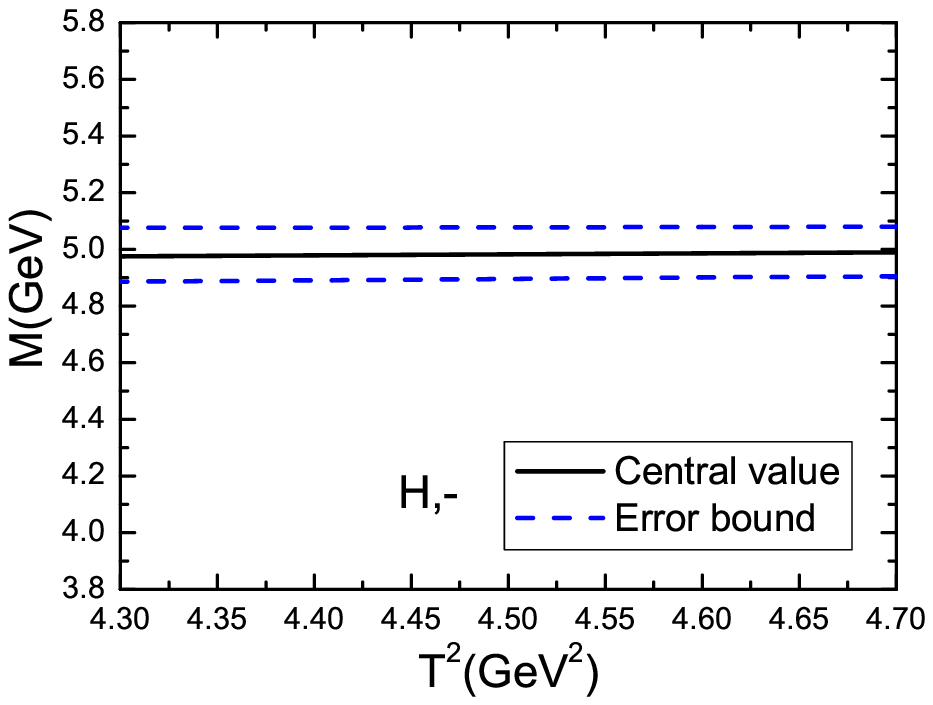}
\includegraphics[totalheight=6cm,width=7cm]{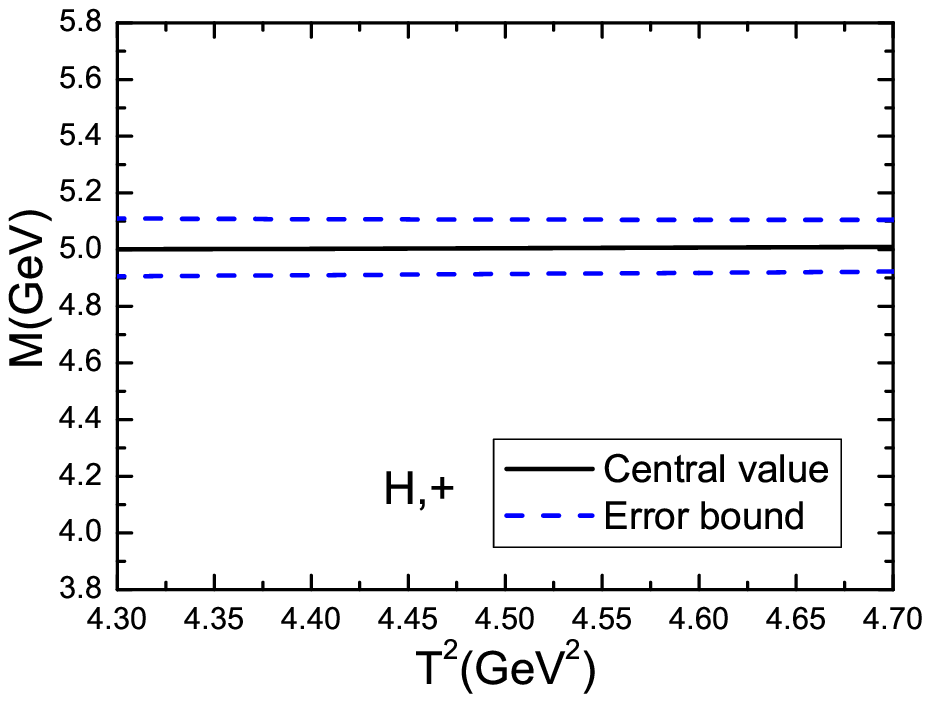}
  \caption{ The masses of the axialvector $cs\bar{c}\bar{s}$ tetraquark states with variations of the  Borel parameters $T^2$, where the horizontal line denotes  the experimental value of the mass $M_{X(4140)}$, the positive sign $+$ (negative sign $-$) denotes the positive charge conjugation (negative charge conjugation). }
\end{figure}

\begin{figure}
\centering
\includegraphics[totalheight=6cm,width=7cm]{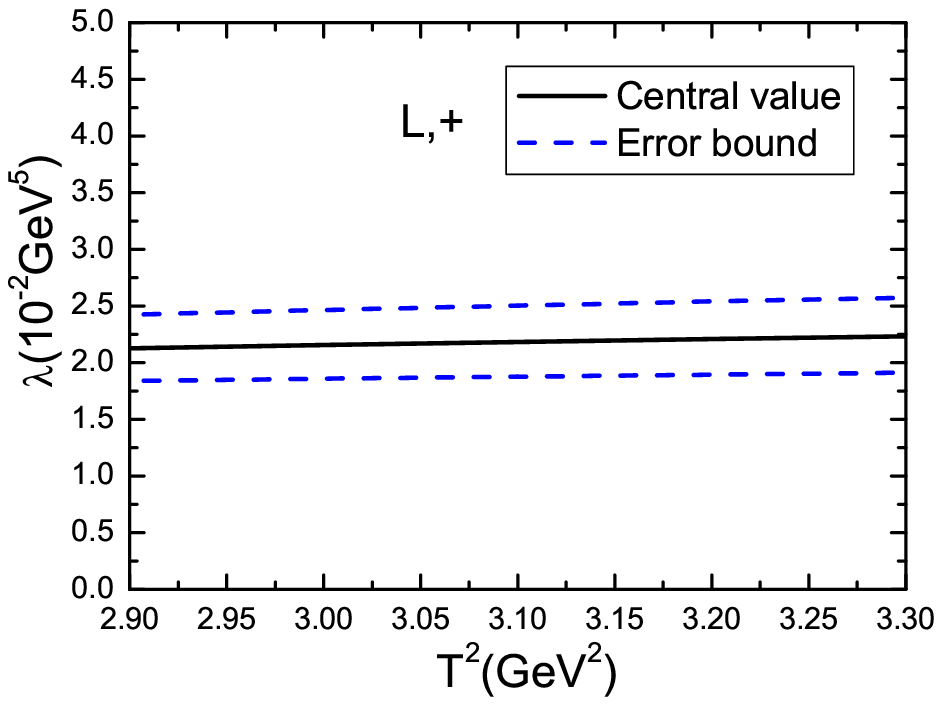}
\includegraphics[totalheight=6cm,width=7cm]{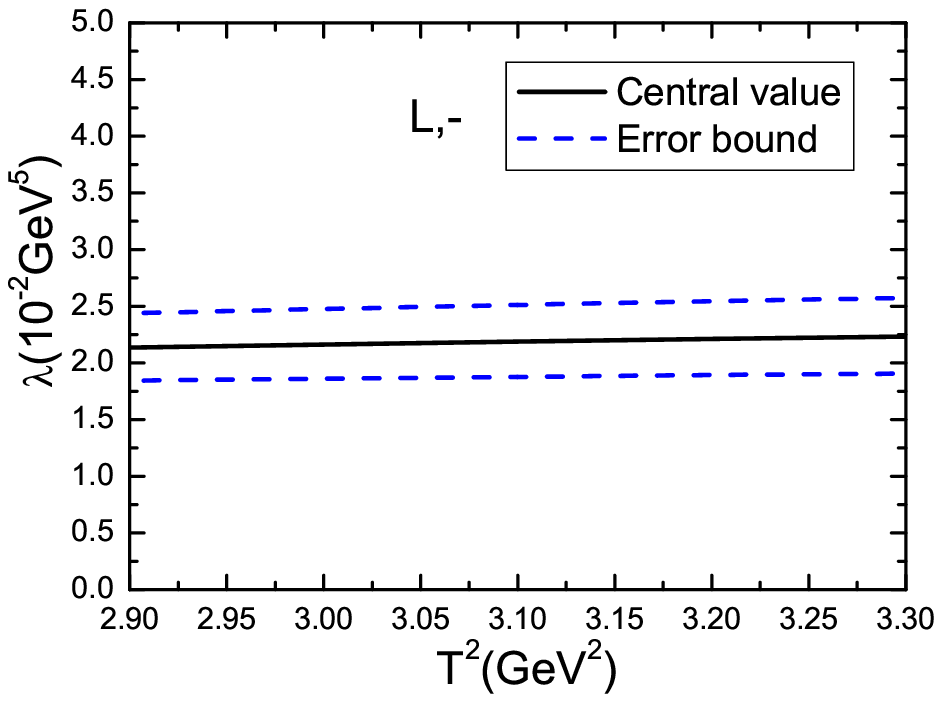}
\includegraphics[totalheight=6cm,width=7cm]{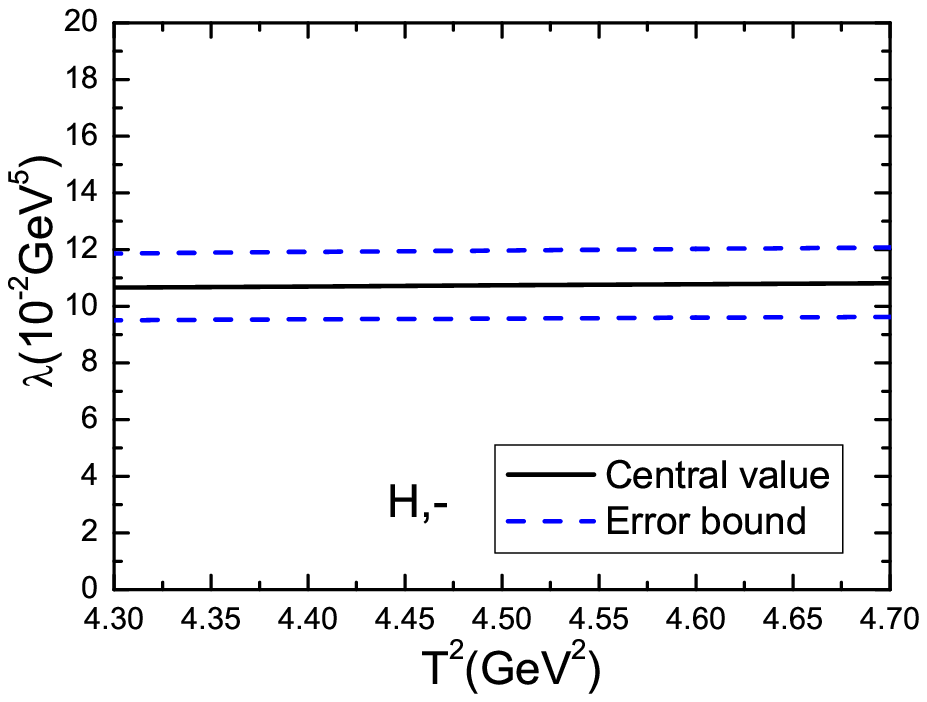}
\includegraphics[totalheight=6cm,width=7cm]{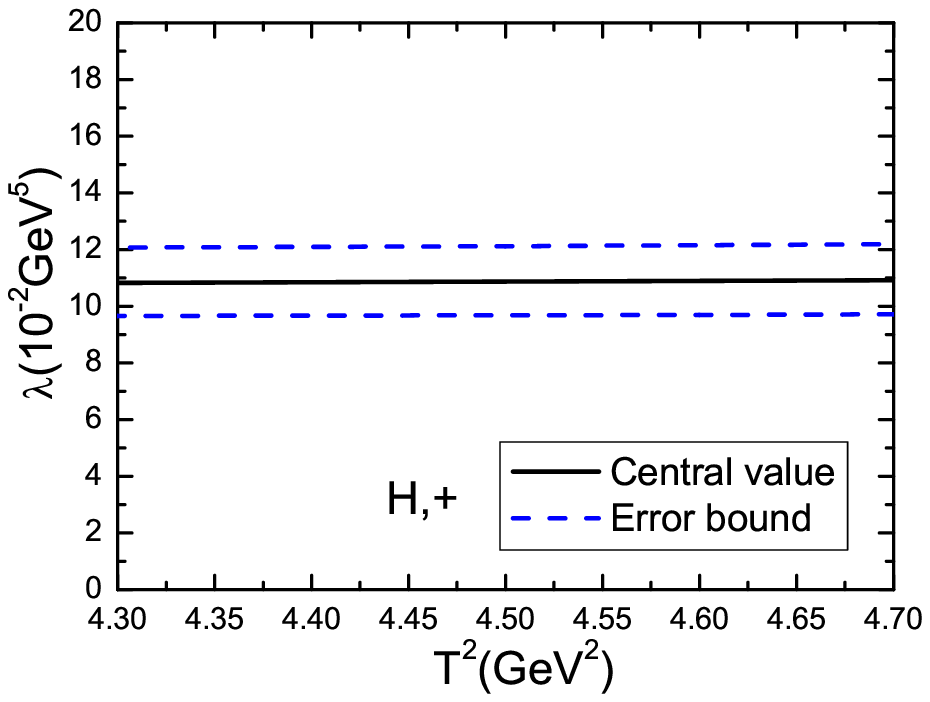}
  \caption{ The pole residues of the axialvector $cs\bar{c}\bar{s}$ tetraquark states with variations of the  Borel parameters $T^2$, the positive sign $+$ (negative sign $-$) denotes the positive charge conjugation (negative charge conjugation). }
\end{figure}

\begin{table}
\begin{center}
\begin{tabular}{|c|c|c|c|c|c|c|c|}\hline\hline
                    & $T^2 (\rm{GeV}^2)$ & $\sqrt{s_0} (\rm{GeV})$ & $\mu(\rm{GeV})$    & pole          & $D_{8}$         & $D_{10}$ \\ \hline
 $X_{L}$ ($1^{++}$) & $2.9-3.3$          & $4.5\pm0.1$             & $1.5$              & $(40-61)\%$   & $-(2-4)\%$      & $\ll 1\%$       \\ \hline
 $X_{L}$ ($1^{+-}$) & $2.9-3.3$          & $4.5\pm0.1$             & $1.5$              & $(39-61)\%$   & $-(4-6)\%$      & $\ll 1\%$    \\ \hline
 $X_{H}$ ($1^{+-}$) & $4.3-4.7$          & $5.5\pm0.1$             & $3.4$              & $(42-58)\%$   & $<1\%$          & $\ll 1\%$    \\ \hline
 $X_{H}$ ($1^{++}$) & $4.3-4.7$          & $5.5\pm0.1$             & $3.4$              & $(41-58)\%$   & $\ll 1\%$       & $\ll 1\%$     \\ \hline
 \hline
\end{tabular}
\end{center}
\caption{ The Borel parameters, continuum threshold parameters, energy scales,  pole contributions, contributions of the vacuum condensates of dimension 8  and 10. }
\end{table}

\begin{table}
\begin{center}
\begin{tabular}{|c|c|c|c|c|c|c|c|}\hline\hline
                      & $M_{X}(\rm{GeV})$         & $\lambda_{X}(10^{-2}\rm{GeV}^5)$ \\ \hline
 $X_{L}$ ($1^{++}$)   & $3.95\pm0.09$             & $2.18\pm0.35$       \\ \hline
 $X_{L}$ ($1^{+-}$)   & $3.97\pm0.09$             & $2.19\pm0.35$    \\ \hline
 $X_{H}$ ($1^{+-}$)   & $4.98\pm0.10$             & $10.7\pm1.2$    \\ \hline
 $X_{H}$ ($1^{++}$)   & $5.00\pm0.10$             & $10.9\pm1.2$    \\ \hline
 \hline
\end{tabular}
\end{center}
\caption{ The masses and pole residues of the axialvector $cs\bar{c}\bar{s}$ tetraquark states. }
\end{table}

Now we perform Fierz re-arrangement to the currents  $J^{L/H,\pm}_\mu$  both in the color space and Dirac-spinor  space, and obtain the following results,
\begin{eqnarray}
J_{L,+}^{\mu} &=&\frac{1}{2\sqrt{2}}\left\{\,\bar{c} \gamma^\mu \gamma_5 c\,\bar{s}  s-\bar{c} c\,\bar{s}\gamma^\mu \gamma_5 s-i\bar{c}i\gamma_5  s\,\bar{s}\gamma^\mu c+i\bar{c} \gamma^\mu s\,\bar{s}i\gamma_5c\right. \nonumber\\
&&\left. - i\bar{c}\gamma_\nu c\, \bar{s}\sigma^{\mu\nu}\gamma_5s+i\bar{c}\sigma^{\mu\nu}\gamma_5 c\, \bar{s}\gamma_\nu s
- i \bar{c}\sigma^{\mu\nu}s\,\bar{s}\gamma_\nu\gamma_5 c+i\bar{c}\gamma_\nu \gamma_5s\, \bar{s}\sigma^{\mu\nu}c   \,\right\} \, , \\
J_{L,-}^{\mu} &=&\frac{1}{2\sqrt{2}}\left\{\,i\bar{c}i\gamma_5 c\,\bar{s}\gamma^\mu s-i\bar{c} \gamma^\mu c\,\bar{s}i\gamma_5 s+\bar{c} s\,\bar{s}\gamma^\mu\gamma_5 c-\bar{c} \gamma^\mu \gamma_5s\,\bar{s}c\right. \nonumber\\
&&\left. - i\bar{c}\gamma_\nu\gamma_5c\, \bar{s}\sigma^{\mu\nu}s+i\bar{c}\sigma^{\mu\nu}c\, \bar{s}\gamma_\nu\gamma_5s
- i \bar{c}\sigma^{\mu\nu}\gamma_5s\,\bar{s}\gamma_\nu c+i\bar{c}\gamma_\nu s\, \bar{s}\sigma^{\mu\nu}\gamma_5c   \,\right\} \, , \\
J_{H,-}^{\mu} &=&\frac{1}{2\sqrt{2}}\left\{\,-i\bar{c} \gamma^\mu  c\,\bar{s}i\gamma_5  s-i\bar{c} i\gamma_5c\,\bar{s}\gamma^\mu  s+i\bar{c}i\gamma_5  s\,\bar{s}\gamma^\mu c+i\bar{c} \gamma^\mu s\,\bar{s}i\gamma_5c\right. \nonumber\\
&&\left. - i\bar{c}\gamma_\nu\gamma_5 c\, \bar{s}\sigma^{\mu\nu} s-i\bar{c}\sigma^{\mu\nu}c\, \bar{s}\gamma_\nu \gamma_5s
+ i \bar{c}\sigma^{\mu\nu}s\,\bar{s}\gamma_\nu \gamma_5c+i\bar{c}\gamma_\nu \gamma_5s\, \bar{s}\sigma^{\mu\nu}c   \,\right\} \, , \\
J_{H,+}^{\mu} &=&\frac{1}{2\sqrt{2}}\left\{\,\bar{c} c\,\bar{s}\gamma^\mu \gamma_5s+\bar{c} \gamma^\mu \gamma_5c\,\bar{s} s-\bar{c} s\,\bar{s}\gamma^\mu\gamma_5 c-\bar{c} \gamma^\mu \gamma_5s\,\bar{s}c\right. \nonumber\\
&&\left. - i\bar{c}\gamma_\nu c\, \bar{s}\sigma^{\mu\nu}\gamma_5s-i\bar{c}\sigma^{\mu\nu}\gamma_5c\, \bar{s}\gamma_\nu s
+ i \bar{c}\sigma^{\mu\nu}\gamma_5s\,\bar{s}\gamma_\nu c+i\bar{c}\gamma_\nu s\, \bar{s}\sigma^{\mu\nu} \gamma_5c   \,\right\} \, ,
\end{eqnarray}
the components such as $\bar{c}i\gamma_5 c\,\bar{s}\gamma^\mu s$, $\bar{c} \gamma^\mu c\,\bar{s}i\gamma_5 s$, $\bar{c}\gamma_\nu c\, \bar{s}\sigma^{\mu\nu}\gamma_5s$, $\bar{c}\sigma^{\mu\nu}\gamma_5c\, \bar{s}\gamma_\nu s$, etc couple  potentially to the molecular states or  meson-meson  pairs.
The physical diquark-antidiquark type tetraquark state can be taken as a special superposition of a series of off-shell molecular states and meson-meson pairs, and embodies  the net effects. The decays to its components (meson-meson pairs) are Okubo-Zweig-Iizuka super-allowed, but the re-arrangements in the color-space are non-trivial. At the phenomenological side of the QCD sum rules, it is not necessary to include the contributions of the molecular states lying nearby the physical tetraquark state explicitly, as their effects are already embodied in the physical tetraquark state.

The two-body strong decays
\begin{eqnarray}
X_{L,+}(1^{++}) &\to& J/\psi \phi \to J/\psi\omega \,\,\, (\phi-\omega\,\,{\rm mixing})\, , \nonumber\\
X_{L,-}(1^{+-}) &\to& \eta_c \phi \, , \, J/\psi \eta \, ,  \\
X_{H,-}(1^{+-}) &\to& \eta_c \phi \, , \, J/\psi \eta \, , \, J/\psi \eta^{\prime}\, , \, D_s^{\pm} D_s^{*\mp}, , \, \chi_{c1}h_1(1380)\, , \,h_{c1}f_1(1420)\,\, , \nonumber\\
X_{H,+}(1^{++}) &\to& J/\psi \phi \, , \,\chi_{c0}f_1(1420) \, , \, D_s^{*\pm} D_s^{*\mp}\, , \, D_{s0}^{*\pm}(2317) D_{s1}^{*\mp}(2460)\, ,
\end{eqnarray}
are Okubo-Zweig-Iizuka   super-allowed. The decay widths of the $X_{L,+}(1^{++})$ and $X_{L,-}(1^{+-})$ are expected to be small due to the small available phase-spaces, while the  decay widths of the $X_{H,+}(1^{++})$ and $X_{H,-}(1^{+-})$ are expected to be large due to the large available phase-spaces.

Now we study the finite width effect on the predicted mass $M_{X_{L,+}}$, which lies in the vicinity of the $M_{X(4140)}$. The   current $J_{\mu}^{L,+}(x)$   couples potentially  to the scattering states  $ J/\psi\omega$, $J/\psi \phi$,  $D_s^{*\pm} D_s^{*\mp}$, $\cdots$, we take into account  the contributions of the  intermediate   meson-loops to the correlation function  $\Pi_{L,+}(p^2)$,
\begin{eqnarray}
\Pi_{L,+}(p^2) &=&-\frac{\widehat{\lambda}_{X_{L,+}}^{2}}{ p^2-\widehat{M}_{X_{L,+}}^2-\Sigma_{J/\psi\omega}(p)-\Sigma_{J/\psi\phi}(p)+\cdots}+\cdots \, ,
\end{eqnarray}
where the $\widehat{\lambda}_{X_{L,+}}$ and $\widehat{M}_{X_{L,+}}$ are bare quantities to absorb the divergences in the self-energies $\Sigma_{J/\psi\omega}(p)$, $\Sigma_{J/\psi\phi}(p)$, $\cdots$.
All the renormalized self-energies  contribute  a finite imaginary part to modify the dispersion relation,
\begin{eqnarray}
\Pi_{L,+}(p^2) &=&-\frac{\lambda_{L,+}^{2}}{ p^2-M_{L,+}^2+i\sqrt{p^2}\Gamma(p^2)}+\cdots \, .
 \end{eqnarray}

We can take into account the finite width effect by the following simple replacement of the hadronic spectral density,
\begin{eqnarray}
\delta \left(s-M^2_{L,+} \right) &\to& \frac{1}{\pi}\frac{\sqrt{s}\,\Gamma_{L,+}(s)}{\left(s-M_{L,+}^2\right)^2+s\,\Gamma_{L,+}^2(s)}\, .
\end{eqnarray}

It is easy to obtain the mass,
\begin{eqnarray}
M_{L,+}^2 &=& \frac{\int_{\Delta^2}^{s^0_{L,+}}ds \, s\, \frac{1}{\pi}\frac{\sqrt{s}\,\Gamma_{L,+}(s)}{\left(s-M_{L,+}^2\right)^2+s\,\Gamma_{L,+}^2(s)}\, \exp\left(-\frac{s}{T^2} \right)}{\int_{\Delta^2}^{s^0_{L,+}}ds \, \frac{1}{\pi}\frac{\sqrt{s}\,\Gamma_{L,+}(s)}{\left(s-M_{L,+}^2\right)^2+s\,\Gamma_{L,+}^2(s)}\, \exp\left(-\frac{s}{T^2} \right)} \, ,
\end{eqnarray}
where the mass $M_{L,+}$ at the right side of Eq.(44) comes from the QCD sum rules in Eq.(29),
$\Gamma_{L,+}(s)=\Gamma_{L,+}$, $\Delta=M_{J/\psi}+M_{\omega}$. The relevant thresholds are $M_{J/\psi}+M_{\phi}=4.11638\,\rm{GeV}$ and $M_{J/\psi}+M_{\omega}=3.87957\,\rm{GeV}$ from the Particle Data Group \cite{PDG}, the decay $X_{L,+} \to J/\psi\phi$ is  kinematically forbidden, the decay $X_{L,+} \to J/\psi\omega$ can take place through  the $\phi-\omega$ mixing.
 The width from the  LHCb collaboration is $\Gamma_{X(4140)} = 83 \pm 21 ^{+21}_{-14} \mbox{ MeV}$  \cite{LHCb-4500-1606.07895,LHCb-4500-1606.07898}, the energy dependence of the small width can be safely neglected. If we assign the $X(4140)$ to be the $X_{L,+}$, then $\Gamma_{L,+}\approx 80\,\rm{MeV}$.
  The numerical result is shown explicitly in Fig.5. From Fig.5, we can see that the predicted mass $M_{L,+}$ increases   monotonously but slowly with the increase of the  finite width $\Gamma_{L,+}$. Now the predicted masses from the QCD sum rules are
\begin{eqnarray}
 M_{L,+}&=&(3.97\pm0.09) \, \rm{GeV} \,\,\, {\rm for}\,\,\, \Gamma_{L,+}=80\,{\rm MeV} \, , \nonumber\\
        &=&(4.00\pm0.09) \, \rm{GeV} \,\,\, {\rm for}\,\,\, \Gamma_{L,+}=200\,{\rm MeV} \, ,
\end{eqnarray}
which are still  smaller than  the experimental value  $M_{X(4140)} = 4146.5 \pm 4.5 ^{+4.6}_{-2.8} \mbox{ MeV}$ from the  LHCb collaboration \cite{LHCb-4500-1606.07895,LHCb-4500-1606.07898}. Moreover, the decay $X_{L,+} \to J/\psi\phi$ is  kinematically forbidden, the total decay width of the $X_{L,+}$ cannot exceed $200\,\rm{MeV}$. The contributions of the  intermediate   meson-loops to the $X_{L,+}$ cannot impair the predictive ability remarkably.

The contributions of the  intermediate   meson-loops to the $X_{L,-}$, $X_{H,-}$, $X_{H,+}$ can be studied analogously. In calculations, we take the thresholds $\Delta=M_{J/\psi}+M_{\eta}=3.64478\,\rm{GeV}$ for the $X_{L,-}$, $X_{H,-}$ and $\Delta=M_{J/\psi}+M_{\omega}=3.87957\,\rm{GeV}$ for the $X_{H,+}$. Moreover,
we take into account of the energy dependence of the finite widths of the  $X_{H,-}$ and $X_{H,+}$,
\begin{eqnarray}
\Gamma_{H,-}(s)&=&\Gamma_{H,-} \frac{M^2_{H,-}}{s}  \sqrt{ \frac{s-(M_{J/\psi}+M_{\eta})^2}{M^2_{H,-}-(M_{J/\psi}+M_{\eta})^2} } \, , \nonumber\\
\Gamma_{H,+}(s)&=&\Gamma_{H,+} \frac{M^2_{H,+}}{s}  \sqrt{ \frac{s-(M_{J/\psi}+M_{\omega})^2}{M^2_{H,+}-(M_{J/\psi}+M_{\omega})^2} } \, .
\end{eqnarray}
The numerical results are also shown in Fig.5.  From the figure, we can see that the predicted mass $M_{L,-}$ decreases   monotonously but very slowly with the increase of the  finite width $\Gamma_{L,-}$, the effect of the finite width $\Gamma_{L,-}$ or the intermediate meson-loops can be neglected safely. However, the predicted masses $M_{H,-}$ and $M_{H,+}$ decrease   monotonously and remarkably  with the increase of the  finite widths $\Gamma_{H,-}$ and $\Gamma_{H,+}$, respectively, as they lie far above the corresponding  thresholds  $\Delta=M_{J/\psi}+M_{\eta}=3.64478\,\rm{GeV}$  and $\Delta=M_{J/\psi}+M_{\omega}=3.87957\,\rm{GeV}$, respectively. For example,
 \begin{eqnarray}
M_{H,- }&=&(4.92\pm0.10) \, \rm{GeV} \,\,\, {\rm for}\,\,\, \Gamma_{H,-}=80\,{\rm MeV} \, , \nonumber\\
        &=&(4.84\pm0.10) \, \rm{GeV} \,\,\, {\rm for}\,\,\, \Gamma_{H,-}=200\,{\rm MeV} \, ,\\
 M_{H,+}&=&(4.96\pm0.10) \, \rm{GeV} \,\,\, {\rm for}\,\,\, \Gamma_{H,+}=80\,{\rm MeV} \, , \nonumber\\
        &=&(4.90\pm0.10) \, \rm{GeV} \,\,\, {\rm for}\,\,\, \Gamma_{H,+}=200\,{\rm MeV} \, .
\end{eqnarray}
 The decays $X_{H,-} \to \eta_c \phi$,  $J/\psi \eta$, $D_s^{\pm} D_s^{*\mp}$ and
$X_{H,+} \to J/\psi \omega$, $J/\psi \phi$,  $D_s^{*\pm} D_s^{*\mp}$ can take place easily, the total decay widths may be large and can modify the predicted masses remarkably, the net effects of the intermediate   meson-loops should be taken into account.

\begin{figure}
\centering
\includegraphics[totalheight=6cm,width=7cm]{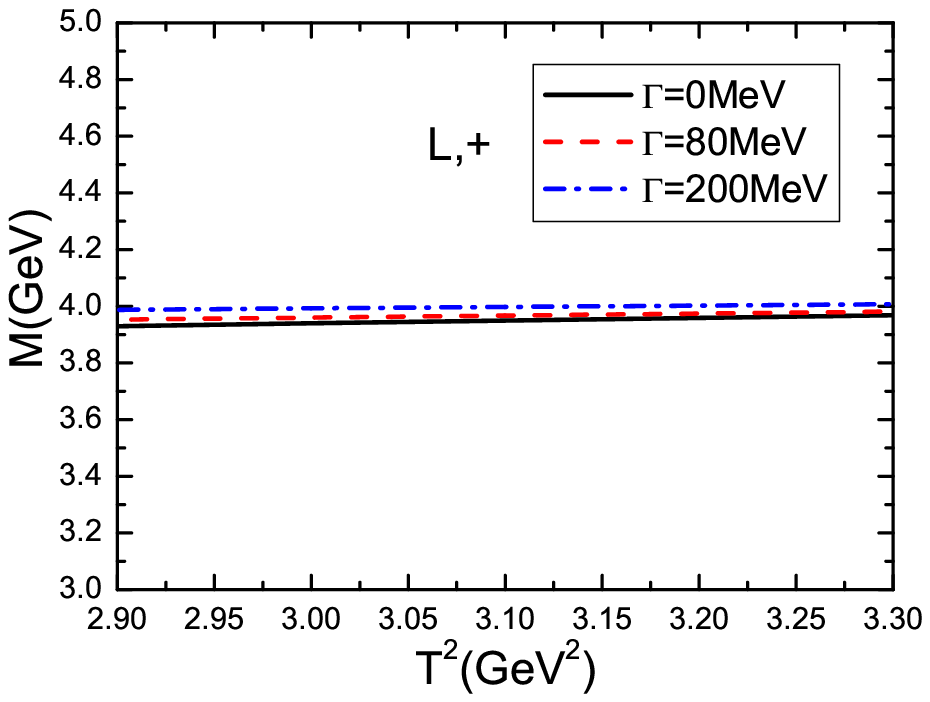}
\includegraphics[totalheight=6cm,width=7cm]{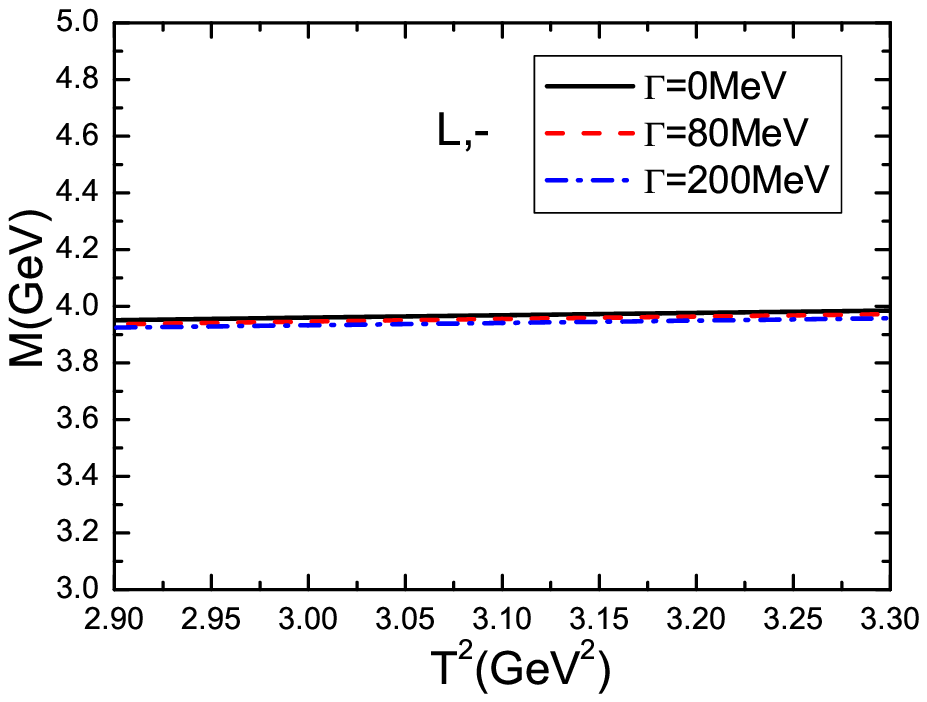}
\includegraphics[totalheight=6cm,width=7cm]{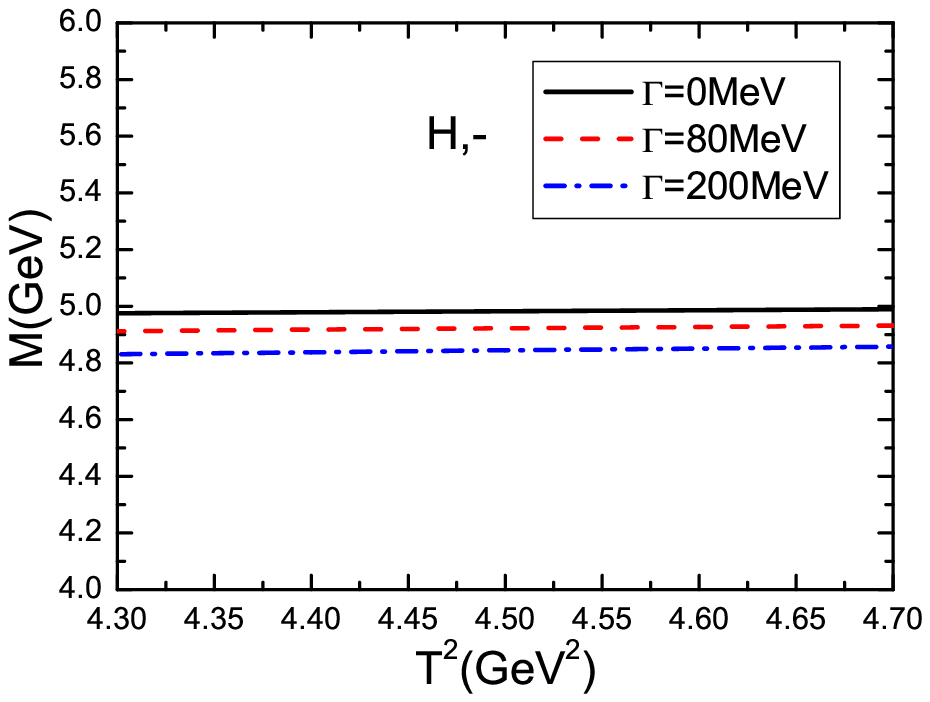}
\includegraphics[totalheight=6cm,width=7cm]{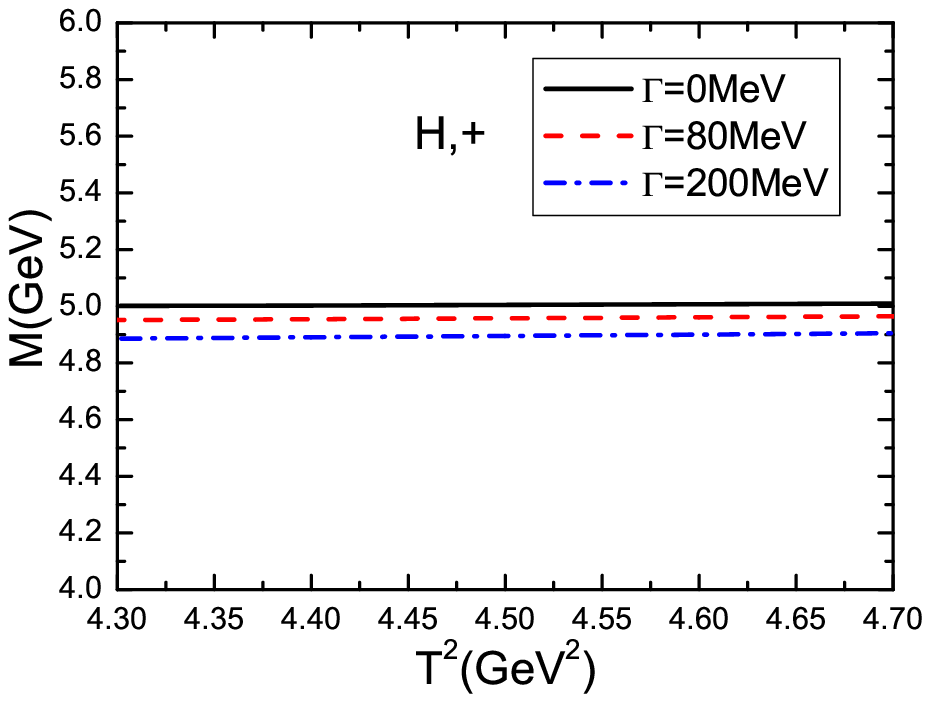}
  \caption{ The masses of the axialvector $cs\bar{c}\bar{s}$ tetraquark states with variations of the  Borel parameters $T^2$ and the finite widths  $\Gamma$, where  the positive sign $+$ (negative sign $-$) denotes the positive charge conjugation (negative charge conjugation). }
\end{figure}

\section{Conclusion}
In this article, we take the $X(4140)$ as the axialvector $cs\bar{c}\bar{s}$ tetraquark state, construct  two diquark-antidiquark type axialvector currents, calculate the contributions of the vacuum condensates up to dimension 10 in the operator product expansion  in a  consistent way, use the empirical energy scale formula to determine the ideal energy scales of the QCD spectral densities, and  study the ground state masses and pole residues with the QCD sum rules. The numerical results $M_{X_{L,+}}=3.95\pm0.09\,\rm{GeV}$ and $M_{X_{H,+}}=5.00\pm0.10\,\rm{GeV}$
disfavor assigning the $X(4140)$ to be the $J^{PC}=1^{++}$ diquark-antidiquark type tetraquark states. Moreover, we obtain the masses of the   $J^{PC}=1^{+-}$ diquark-antidiquark type $cs\bar{c}\bar{s}$ tetraquark states as a byproduct. The present predictions of the masses of the axialvector $cs\bar{c}\bar{s}$ tetraquark states can be confronted to the experimental data in the future.

\section*{Acknowledgements}
This  work is supported by National Natural Science Foundation,
Grant Numbers 11375063,  and Natural Science Foundation of Hebei province, Grant Number A2014502017.

\end{document}